 \documentclass[aps, pre,amsmath,amssymb,twocolumn,floatfix,showpacs]{revtex4} 
\usepackage[usenames,dvipsnames]{color} 
\usepackage[pdftex]{graphicx} 
\usepackage{epsfig} 
\usepackage{pgf}
\usepackage{mathtools}

\graphicspath{{figs/}}

     {\begin{list}{}%
             {\setlength{\leftmargin}{#1}}%
             \item[]%
     }
     {\end{list}}

\def\i{\mathbf{i}}
\def\j{\mathbf{j}}
\def\e{\varepsilon}

\begin{document}

\title{Modelling the evaporation of nanoparticle
suspensions from heterogeneous surfaces}
\author{C. Chalmers, R. Smith and A.J. Archer}
\affiliation{Department of Mathematical Sciences,
  Loughborough University,
  Loughborough LE11 3TU,
  UK
}

\begin{abstract}
  We present a Monte Carlo (MC) grid-based model for the drying
  of drops of a nanoparticle suspension  upon a heterogeneous surface.
  The model consists of a generalised lattice-gas in which the
  interaction parameters in the Hamiltonian can be varied to
  model different properties of the materials involved. We show how to
   choose correctly the interactions,  to minimise the effects of the underlying grid so that hemispherical droplets
  form. We also include
  the effects of surface roughness to examine the effects of contact-line
  pinning on the dynamics. When there is a `lid' above the system, which
  prevents evaporation, equilibrium drops form on the surface, which we
  use to determine the contact angle and how it varies as the parameters of
  the model are changed. This enables us to relate the interaction
  parameters to the materials used in applications.
  The model has also been applied to drying on heterogeneous surfaces,
  in particular to the case where the suspension is deposited on a surface
  consisting of a pair of hydrophilic conducting metal surfaces that are
  either side of a band of hydrophobic insulating polymer.
  This situation occurs when using inkjet printing to
  manufacture electrical connections between the metallic parts of the
  surface. The process is not always without problems, since the liquid
  can dewet from the hydrophobic part of the surface, breaking the
  bridge before the drying process is complete. The MC model reproduces
  the observed dewetting, allowing the parameters to be varied so
  that the conditions for the best connection can be established. We
  show that if the hydrophobic portion of the surface is located at a
  step below the height of the neighbouring metal, the chance of
  dewetting of the liquid during the drying process is significantly
  reduced.
\end{abstract}

\maketitle
\epsfclipon 

\section{Introduction}\label{introduction}

How ink or paint dries, i.e.\ how liquid droplets containing
nanoparticles deposited on a surface evolve in time as the liquid
evaporates has significant relevance in modern manufacturing. Inkjet
deposition is increasingly used during the manufacture of functional
nano-structured materials. An innovative recent example is
the application described in \cite{walls} which uses inkjet printing as
part of a new method for constructing solar panels. This includes making
the electrical interconnections after the various different layers that
form a solar cell have been laid down on the glass substrate and then
scribed using laser ablation~\cite{walls}. The benefits of using inkjet
printing include reduced costs, wastage and potentially improved
performance. Several parts of the structure are inkjet printed. An
insulating polymer layer can be formed by depositing an ink that
consists of a polymer solution. Another ink consists of a suspension of
conducting metal nanoparticles. As the solvent evaporates, the goal is
for the remaining nanoparticles to form an electrically conducting
connection over the surface.

\begin{figure}
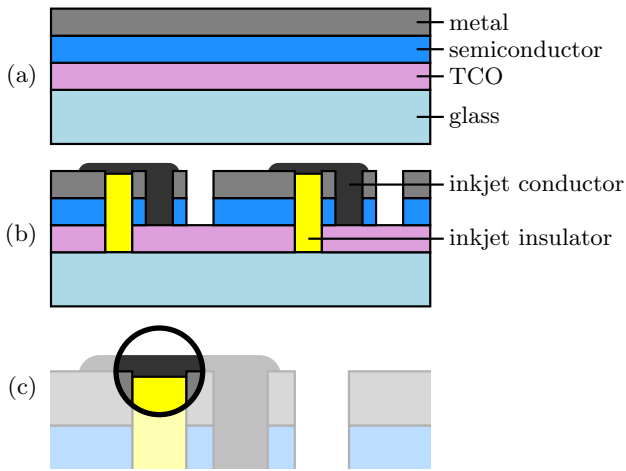

\begin{pgfpicture}
\pgfpathrectangle{\pgfpointorigin}{\pgfqpoint{245.0000bp}{173.0000bp}}
\pgfusepath{use as bounding box}
\begin{pgfscope}
\pgfpathqmoveto{165.3750bp}{0.0000bp}
\pgfpathqlineto{165.3750bp}{61.2500bp}
\pgfpathqlineto{22.4583bp}{61.2500bp}
\pgfpathqlineto{22.4583bp}{0.0000bp}
\pgfpathqlineto{165.3750bp}{0.0000bp}
\pgfpathclose
\pgfusepathqclip
\begin{pgfscope}
\pgfpathqmoveto{165.3750bp}{0.0000bp}
\pgfpathqlineto{165.3750bp}{61.2500bp}
\pgfpathqlineto{22.4583bp}{61.2500bp}
\pgfpathqlineto{22.4583bp}{0.0000bp}
\pgfpathqlineto{165.3750bp}{0.0000bp}
\pgfpathclose
\pgfusepathqclip
\begin{pgfscope}
\definecolor{fc}{rgb}{0.2000,0.2000,0.2000}
\pgfsetfillcolor{fc}
\pgfpathqmoveto{252.1458bp}{-6.1250bp}
\pgfpathqlineto{252.1458bp}{34.7083bp}
\pgfpathqcurveto{252.1458bp}{39.2187bp}{248.4895bp}{42.8750bp}{243.9792bp}{42.8750bp}
\pgfpathqlineto{183.7500bp}{42.8750bp}
\pgfpathqcurveto{179.2397bp}{42.8750bp}{175.5833bp}{39.2187bp}{175.5833bp}{34.7083bp}
\pgfpathqlineto{175.5833bp}{-6.1250bp}
\pgfpathqcurveto{175.5833bp}{-10.6353bp}{179.2397bp}{-14.2917bp}{183.7500bp}{-14.2917bp}
\pgfpathqlineto{243.9792bp}{-14.2917bp}
\pgfpathqcurveto{248.4895bp}{-14.2917bp}{252.1458bp}{-10.6353bp}{252.1458bp}{-6.1250bp}
\pgfpathclose
\pgfusepathqfill
\end{pgfscope}
\end{pgfscope}
\begin{pgfscope}
\pgfpathqmoveto{165.3750bp}{0.0000bp}
\pgfpathqlineto{165.3750bp}{61.2500bp}
\pgfpathqlineto{22.4583bp}{61.2500bp}
\pgfpathqlineto{22.4583bp}{0.0000bp}
\pgfpathqlineto{165.3750bp}{0.0000bp}
\pgfpathclose
\pgfusepathqclip
\begin{pgfscope}
\definecolor{fc}{rgb}{1.0000,1.0000,0.0000}
\pgfsetfillcolor{fc}
\pgfsetlinewidth{0.8248bp}
\definecolor{sc}{rgb}{0.0000,0.0000,0.0000}
\pgfsetstrokecolor{sc}
\pgfsetmiterjoin
\pgfsetbuttcap
\pgfpathqmoveto{216.4167bp}{-24.5000bp}
\pgfpathqlineto{216.4167bp}{34.7083bp}
\pgfpathqlineto{196.0000bp}{34.7083bp}
\pgfpathqlineto{196.0000bp}{-24.5000bp}
\pgfpathqlineto{216.4167bp}{-24.5000bp}
\pgfpathclose
\pgfusepathqfillstroke
\end{pgfscope}
\end{pgfscope}
\begin{pgfscope}
\pgfpathqmoveto{165.3750bp}{0.0000bp}
\pgfpathqlineto{165.3750bp}{61.2500bp}
\pgfpathqlineto{22.4583bp}{61.2500bp}
\pgfpathqlineto{22.4583bp}{0.0000bp}
\pgfpathqlineto{165.3750bp}{0.0000bp}
\pgfpathclose
\pgfusepathqclip
\begin{pgfscope}
\definecolor{fc}{rgb}{0.2000,0.2000,0.2000}
\pgfsetfillcolor{fc}
\pgfpathqmoveto{109.2292bp}{-6.1250bp}
\pgfpathqlineto{109.2292bp}{34.7083bp}
\pgfpathqcurveto{109.2292bp}{39.2187bp}{105.5728bp}{42.8750bp}{101.0625bp}{42.8750bp}
\pgfpathqlineto{40.8333bp}{42.8750bp}
\pgfpathqcurveto{36.3230bp}{42.8750bp}{32.6667bp}{39.2187bp}{32.6667bp}{34.7083bp}
\pgfpathqlineto{32.6667bp}{-6.1250bp}
\pgfpathqcurveto{32.6667bp}{-10.6353bp}{36.3230bp}{-14.2917bp}{40.8333bp}{-14.2917bp}
\pgfpathqlineto{101.0625bp}{-14.2917bp}
\pgfpathqcurveto{105.5728bp}{-14.2917bp}{109.2292bp}{-10.6353bp}{109.2292bp}{-6.1250bp}
\pgfpathclose
\pgfusepathqfill
\end{pgfscope}
\end{pgfscope}
\begin{pgfscope}
\pgfpathqmoveto{165.3750bp}{0.0000bp}
\pgfpathqlineto{165.3750bp}{61.2500bp}
\pgfpathqlineto{22.4583bp}{61.2500bp}
\pgfpathqlineto{22.4583bp}{0.0000bp}
\pgfpathqlineto{165.3750bp}{0.0000bp}
\pgfpathclose
\pgfusepathqclip
\begin{pgfscope}
\definecolor{fc}{rgb}{1.0000,1.0000,0.0000}
\pgfsetfillcolor{fc}
\pgfsetlinewidth{0.8248bp}
\definecolor{sc}{rgb}{0.0000,0.0000,0.0000}
\pgfsetstrokecolor{sc}
\pgfsetmiterjoin
\pgfsetbuttcap
\pgfpathqmoveto{73.5000bp}{-24.5000bp}
\pgfpathqlineto{73.5000bp}{34.7083bp}
\pgfpathqlineto{53.0833bp}{34.7083bp}
\pgfpathqlineto{53.0833bp}{-24.5000bp}
\pgfpathqlineto{73.5000bp}{-24.5000bp}
\pgfpathclose
\pgfusepathqfillstroke
\end{pgfscope}
\end{pgfscope}
\begin{pgfscope}
\pgfpathqmoveto{165.3750bp}{0.0000bp}
\pgfpathqlineto{165.3750bp}{61.2500bp}
\pgfpathqlineto{22.4583bp}{61.2500bp}
\pgfpathqlineto{22.4583bp}{0.0000bp}
\pgfpathqlineto{165.3750bp}{0.0000bp}
\pgfpathclose
\pgfusepathqclip
\begin{pgfscope}
\pgfpathqmoveto{165.3750bp}{0.0000bp}
\pgfpathqlineto{165.3750bp}{61.2500bp}
\pgfpathqlineto{22.4583bp}{61.2500bp}
\pgfpathqlineto{22.4583bp}{0.0000bp}
\pgfpathqlineto{165.3750bp}{0.0000bp}
\pgfpathclose
\pgfusepathqclip
\begin{pgfscope}
\pgfpathqmoveto{165.3750bp}{0.0000bp}
\pgfpathqlineto{165.3750bp}{61.2500bp}
\pgfpathqlineto{22.4583bp}{61.2500bp}
\pgfpathqlineto{22.4583bp}{0.0000bp}
\pgfpathqlineto{165.3750bp}{0.0000bp}
\pgfpathclose
\pgfusepathqclip
\begin{pgfscope}
\definecolor{fc}{rgb}{0.6784,0.8471,0.9020}
\pgfsetfillcolor{fc}
\pgfsetlinewidth{0.8248bp}
\definecolor{sc}{rgb}{0.0000,0.0000,0.0000}
\pgfsetstrokecolor{sc}
\pgfsetmiterjoin
\pgfsetbuttcap
\pgfpathqmoveto{298.0833bp}{-65.3333bp}
\pgfpathqlineto{298.0833bp}{-24.5000bp}
\pgfpathqlineto{12.2500bp}{-24.5000bp}
\pgfpathqlineto{12.2500bp}{-65.3333bp}
\pgfpathqlineto{298.0833bp}{-65.3333bp}
\pgfpathclose
\pgfusepathqfillstroke
\end{pgfscope}
\end{pgfscope}
\begin{pgfscope}
\pgfpathqmoveto{165.3750bp}{0.0000bp}
\pgfpathqlineto{165.3750bp}{61.2500bp}
\pgfpathqlineto{22.4583bp}{61.2500bp}
\pgfpathqlineto{22.4583bp}{0.0000bp}
\pgfpathqlineto{165.3750bp}{0.0000bp}
\pgfpathclose
\pgfusepathqclip
\begin{pgfscope}
\pgfpathqmoveto{165.3750bp}{0.0000bp}
\pgfpathqlineto{165.3750bp}{61.2500bp}
\pgfpathqlineto{22.4583bp}{61.2500bp}
\pgfpathqlineto{22.4583bp}{0.0000bp}
\pgfpathqlineto{165.3750bp}{0.0000bp}
\pgfpathclose
\pgfusepathqclip
\begin{pgfscope}
\definecolor{fc}{rgb}{0.8667,0.6275,0.8667}
\pgfsetfillcolor{fc}
\pgfsetlinewidth{0.8248bp}
\definecolor{sc}{rgb}{0.0000,0.0000,0.0000}
\pgfsetstrokecolor{sc}
\pgfsetmiterjoin
\pgfsetbuttcap
\pgfpathqmoveto{298.0833bp}{-24.5000bp}
\pgfpathqlineto{298.0833bp}{-4.0833bp}
\pgfpathqlineto{216.4167bp}{-4.0833bp}
\pgfpathqlineto{216.4167bp}{-24.5000bp}
\pgfpathqlineto{298.0833bp}{-24.5000bp}
\pgfpathclose
\pgfusepathqfillstroke
\end{pgfscope}
\end{pgfscope}
\begin{pgfscope}
\pgfpathqmoveto{165.3750bp}{0.0000bp}
\pgfpathqlineto{165.3750bp}{61.2500bp}
\pgfpathqlineto{22.4583bp}{61.2500bp}
\pgfpathqlineto{22.4583bp}{0.0000bp}
\pgfpathqlineto{165.3750bp}{0.0000bp}
\pgfpathclose
\pgfusepathqclip
\begin{pgfscope}
\definecolor{fc}{rgb}{0.8667,0.6275,0.8667}
\pgfsetfillcolor{fc}
\pgfsetlinewidth{0.8248bp}
\definecolor{sc}{rgb}{0.0000,0.0000,0.0000}
\pgfsetstrokecolor{sc}
\pgfsetmiterjoin
\pgfsetbuttcap
\pgfpathqmoveto{196.0000bp}{-24.5000bp}
\pgfpathqlineto{196.0000bp}{-4.0833bp}
\pgfpathqlineto{73.5000bp}{-4.0833bp}
\pgfpathqlineto{73.5000bp}{-24.5000bp}
\pgfpathqlineto{196.0000bp}{-24.5000bp}
\pgfpathclose
\pgfusepathqfillstroke
\end{pgfscope}
\end{pgfscope}
\begin{pgfscope}
\definecolor{fc}{rgb}{0.8667,0.6275,0.8667}
\pgfsetfillcolor{fc}
\pgfsetlinewidth{0.8248bp}
\definecolor{sc}{rgb}{0.0000,0.0000,0.0000}
\pgfsetstrokecolor{sc}
\pgfsetmiterjoin
\pgfsetbuttcap
\pgfpathqmoveto{53.0833bp}{-24.5000bp}
\pgfpathqlineto{53.0833bp}{-4.0833bp}
\pgfpathqlineto{12.2500bp}{-4.0833bp}
\pgfpathqlineto{12.2500bp}{-24.5000bp}
\pgfpathqlineto{53.0833bp}{-24.5000bp}
\pgfpathclose
\pgfusepathqfillstroke
\end{pgfscope}
\end{pgfscope}
\end{pgfscope}
\begin{pgfscope}
\pgfpathqmoveto{165.3750bp}{0.0000bp}
\pgfpathqlineto{165.3750bp}{61.2500bp}
\pgfpathqlineto{22.4583bp}{61.2500bp}
\pgfpathqlineto{22.4583bp}{0.0000bp}
\pgfpathqlineto{165.3750bp}{0.0000bp}
\pgfpathclose
\pgfusepathqclip
\begin{pgfscope}
\pgfpathqmoveto{165.3750bp}{0.0000bp}
\pgfpathqlineto{165.3750bp}{61.2500bp}
\pgfpathqlineto{22.4583bp}{61.2500bp}
\pgfpathqlineto{22.4583bp}{0.0000bp}
\pgfpathqlineto{165.3750bp}{0.0000bp}
\pgfpathclose
\pgfusepathqclip
\begin{pgfscope}
\pgfpathqmoveto{165.3750bp}{0.0000bp}
\pgfpathqlineto{165.3750bp}{61.2500bp}
\pgfpathqlineto{22.4583bp}{61.2500bp}
\pgfpathqlineto{22.4583bp}{0.0000bp}
\pgfpathqlineto{165.3750bp}{0.0000bp}
\pgfpathclose
\pgfusepathqclip
\begin{pgfscope}
\definecolor{fc}{rgb}{0.1176,0.5647,1.0000}
\pgfsetfillcolor{fc}
\pgfsetlinewidth{0.8248bp}
\definecolor{sc}{rgb}{0.0000,0.0000,0.0000}
\pgfsetstrokecolor{sc}
\pgfsetmiterjoin
\pgfsetbuttcap
\pgfpathqmoveto{298.0833bp}{-4.0833bp}
\pgfpathqlineto{298.0833bp}{16.3333bp}
\pgfpathqlineto{277.6667bp}{16.3333bp}
\pgfpathqlineto{277.6667bp}{-4.0833bp}
\pgfpathqlineto{298.0833bp}{-4.0833bp}
\pgfpathclose
\pgfusepathqfillstroke
\end{pgfscope}
\end{pgfscope}
\begin{pgfscope}
\pgfpathqmoveto{165.3750bp}{0.0000bp}
\pgfpathqlineto{165.3750bp}{61.2500bp}
\pgfpathqlineto{22.4583bp}{61.2500bp}
\pgfpathqlineto{22.4583bp}{0.0000bp}
\pgfpathqlineto{165.3750bp}{0.0000bp}
\pgfpathclose
\pgfusepathqclip
\begin{pgfscope}
\definecolor{fc}{rgb}{0.1176,0.5647,1.0000}
\pgfsetfillcolor{fc}
\pgfsetlinewidth{0.8248bp}
\definecolor{sc}{rgb}{0.0000,0.0000,0.0000}
\pgfsetstrokecolor{sc}
\pgfsetmiterjoin
\pgfsetbuttcap
\pgfpathqmoveto{257.2500bp}{-4.0833bp}
\pgfpathqlineto{257.2500bp}{16.3333bp}
\pgfpathqlineto{247.0417bp}{16.3333bp}
\pgfpathqlineto{247.0417bp}{-4.0833bp}
\pgfpathqlineto{257.2500bp}{-4.0833bp}
\pgfpathclose
\pgfusepathqfillstroke
\end{pgfscope}
\end{pgfscope}
\begin{pgfscope}
\definecolor{fc}{rgb}{0.1176,0.5647,1.0000}
\pgfsetfillcolor{fc}
\pgfsetlinewidth{0.8248bp}
\definecolor{sc}{rgb}{0.0000,0.0000,0.0000}
\pgfsetstrokecolor{sc}
\pgfsetmiterjoin
\pgfsetbuttcap
\pgfpathqmoveto{226.6250bp}{-4.0833bp}
\pgfpathqlineto{226.6250bp}{16.3333bp}
\pgfpathqlineto{216.4167bp}{16.3333bp}
\pgfpathqlineto{216.4167bp}{-4.0833bp}
\pgfpathqlineto{226.6250bp}{-4.0833bp}
\pgfpathclose
\pgfusepathqfillstroke
\end{pgfscope}
\end{pgfscope}
\begin{pgfscope}
\pgfpathqmoveto{165.3750bp}{0.0000bp}
\pgfpathqlineto{165.3750bp}{61.2500bp}
\pgfpathqlineto{22.4583bp}{61.2500bp}
\pgfpathqlineto{22.4583bp}{0.0000bp}
\pgfpathqlineto{165.3750bp}{0.0000bp}
\pgfpathclose
\pgfusepathqclip
\begin{pgfscope}
\pgfpathqmoveto{165.3750bp}{0.0000bp}
\pgfpathqlineto{165.3750bp}{61.2500bp}
\pgfpathqlineto{22.4583bp}{61.2500bp}
\pgfpathqlineto{22.4583bp}{0.0000bp}
\pgfpathqlineto{165.3750bp}{0.0000bp}
\pgfpathclose
\pgfusepathqclip
\begin{pgfscope}
\definecolor{fc}{rgb}{0.1176,0.5647,1.0000}
\pgfsetfillcolor{fc}
\pgfsetlinewidth{0.8248bp}
\definecolor{sc}{rgb}{0.0000,0.0000,0.0000}
\pgfsetstrokecolor{sc}
\pgfsetmiterjoin
\pgfsetbuttcap
\pgfpathqmoveto{196.0000bp}{-4.0833bp}
\pgfpathqlineto{196.0000bp}{16.3333bp}
\pgfpathqlineto{134.7500bp}{16.3333bp}
\pgfpathqlineto{134.7500bp}{-4.0833bp}
\pgfpathqlineto{196.0000bp}{-4.0833bp}
\pgfpathclose
\pgfusepathqfillstroke
\end{pgfscope}
\end{pgfscope}
\begin{pgfscope}
\definecolor{fc}{rgb}{0.1176,0.5647,1.0000}
\pgfsetfillcolor{fc}
\pgfsetlinewidth{0.8248bp}
\definecolor{sc}{rgb}{0.0000,0.0000,0.0000}
\pgfsetstrokecolor{sc}
\pgfsetmiterjoin
\pgfsetbuttcap
\pgfpathqmoveto{114.3333bp}{-4.0833bp}
\pgfpathqlineto{114.3333bp}{16.3333bp}
\pgfpathqlineto{104.1250bp}{16.3333bp}
\pgfpathqlineto{104.1250bp}{-4.0833bp}
\pgfpathqlineto{114.3333bp}{-4.0833bp}
\pgfpathclose
\pgfusepathqfillstroke
\end{pgfscope}
\end{pgfscope}
\begin{pgfscope}
\pgfpathqmoveto{165.3750bp}{0.0000bp}
\pgfpathqlineto{165.3750bp}{61.2500bp}
\pgfpathqlineto{22.4583bp}{61.2500bp}
\pgfpathqlineto{22.4583bp}{0.0000bp}
\pgfpathqlineto{165.3750bp}{0.0000bp}
\pgfpathclose
\pgfusepathqclip
\begin{pgfscope}
\definecolor{fc}{rgb}{0.1176,0.5647,1.0000}
\pgfsetfillcolor{fc}
\pgfsetlinewidth{0.8248bp}
\definecolor{sc}{rgb}{0.0000,0.0000,0.0000}
\pgfsetstrokecolor{sc}
\pgfsetmiterjoin
\pgfsetbuttcap
\pgfpathqmoveto{83.7083bp}{-4.0833bp}
\pgfpathqlineto{83.7083bp}{16.3333bp}
\pgfpathqlineto{73.5000bp}{16.3333bp}
\pgfpathqlineto{73.5000bp}{-4.0833bp}
\pgfpathqlineto{83.7083bp}{-4.0833bp}
\pgfpathclose
\pgfusepathqfillstroke
\end{pgfscope}
\end{pgfscope}
\begin{pgfscope}
\definecolor{fc}{rgb}{0.1176,0.5647,1.0000}
\pgfsetfillcolor{fc}
\pgfsetlinewidth{0.8248bp}
\definecolor{sc}{rgb}{0.0000,0.0000,0.0000}
\pgfsetstrokecolor{sc}
\pgfsetmiterjoin
\pgfsetbuttcap
\pgfpathqmoveto{53.0833bp}{-4.0833bp}
\pgfpathqlineto{53.0833bp}{16.3333bp}
\pgfpathqlineto{12.2500bp}{16.3333bp}
\pgfpathqlineto{12.2500bp}{-4.0833bp}
\pgfpathqlineto{53.0833bp}{-4.0833bp}
\pgfpathclose
\pgfusepathqfillstroke
\end{pgfscope}
\end{pgfscope}
\begin{pgfscope}
\pgfpathqmoveto{165.3750bp}{0.0000bp}
\pgfpathqlineto{165.3750bp}{61.2500bp}
\pgfpathqlineto{22.4583bp}{61.2500bp}
\pgfpathqlineto{22.4583bp}{0.0000bp}
\pgfpathqlineto{165.3750bp}{0.0000bp}
\pgfpathclose
\pgfusepathqclip
\begin{pgfscope}
\pgfpathqmoveto{165.3750bp}{0.0000bp}
\pgfpathqlineto{165.3750bp}{61.2500bp}
\pgfpathqlineto{22.4583bp}{61.2500bp}
\pgfpathqlineto{22.4583bp}{0.0000bp}
\pgfpathqlineto{165.3750bp}{0.0000bp}
\pgfpathclose
\pgfusepathqclip
\begin{pgfscope}
\pgfpathqmoveto{165.3750bp}{0.0000bp}
\pgfpathqlineto{165.3750bp}{61.2500bp}
\pgfpathqlineto{22.4583bp}{61.2500bp}
\pgfpathqlineto{22.4583bp}{0.0000bp}
\pgfpathqlineto{165.3750bp}{0.0000bp}
\pgfpathclose
\pgfusepathqclip
\begin{pgfscope}
\definecolor{fc}{rgb}{0.5000,0.5000,0.5000}
\pgfsetfillcolor{fc}
\pgfsetlinewidth{0.8248bp}
\definecolor{sc}{rgb}{0.0000,0.0000,0.0000}
\pgfsetstrokecolor{sc}
\pgfsetmiterjoin
\pgfsetbuttcap
\pgfpathqmoveto{298.0833bp}{16.3333bp}
\pgfpathqlineto{298.0833bp}{36.7500bp}
\pgfpathqlineto{277.6667bp}{36.7500bp}
\pgfpathqlineto{277.6667bp}{16.3333bp}
\pgfpathqlineto{298.0833bp}{16.3333bp}
\pgfpathclose
\pgfusepathqfillstroke
\end{pgfscope}
\end{pgfscope}
\begin{pgfscope}
\pgfpathqmoveto{165.3750bp}{0.0000bp}
\pgfpathqlineto{165.3750bp}{61.2500bp}
\pgfpathqlineto{22.4583bp}{61.2500bp}
\pgfpathqlineto{22.4583bp}{0.0000bp}
\pgfpathqlineto{165.3750bp}{0.0000bp}
\pgfpathclose
\pgfusepathqclip
\begin{pgfscope}
\definecolor{fc}{rgb}{0.5000,0.5000,0.5000}
\pgfsetfillcolor{fc}
\pgfsetlinewidth{0.8248bp}
\definecolor{sc}{rgb}{0.0000,0.0000,0.0000}
\pgfsetstrokecolor{sc}
\pgfsetmiterjoin
\pgfsetbuttcap
\pgfpathqmoveto{257.2500bp}{16.3333bp}
\pgfpathqlineto{257.2500bp}{36.7500bp}
\pgfpathqlineto{247.0417bp}{36.7500bp}
\pgfpathqlineto{247.0417bp}{16.3333bp}
\pgfpathqlineto{257.2500bp}{16.3333bp}
\pgfpathclose
\pgfusepathqfillstroke
\end{pgfscope}
\end{pgfscope}
\begin{pgfscope}
\definecolor{fc}{rgb}{0.5000,0.5000,0.5000}
\pgfsetfillcolor{fc}
\pgfsetlinewidth{0.8248bp}
\definecolor{sc}{rgb}{0.0000,0.0000,0.0000}
\pgfsetstrokecolor{sc}
\pgfsetmiterjoin
\pgfsetbuttcap
\pgfpathqmoveto{226.6250bp}{16.3333bp}
\pgfpathqlineto{226.6250bp}{36.7500bp}
\pgfpathqlineto{216.4167bp}{36.7500bp}
\pgfpathqlineto{216.4167bp}{16.3333bp}
\pgfpathqlineto{226.6250bp}{16.3333bp}
\pgfpathclose
\pgfusepathqfillstroke
\end{pgfscope}
\end{pgfscope}
\begin{pgfscope}
\pgfpathqmoveto{165.3750bp}{0.0000bp}
\pgfpathqlineto{165.3750bp}{61.2500bp}
\pgfpathqlineto{22.4583bp}{61.2500bp}
\pgfpathqlineto{22.4583bp}{0.0000bp}
\pgfpathqlineto{165.3750bp}{0.0000bp}
\pgfpathclose
\pgfusepathqclip
\begin{pgfscope}
\pgfpathqmoveto{165.3750bp}{0.0000bp}
\pgfpathqlineto{165.3750bp}{61.2500bp}
\pgfpathqlineto{22.4583bp}{61.2500bp}
\pgfpathqlineto{22.4583bp}{0.0000bp}
\pgfpathqlineto{165.3750bp}{0.0000bp}
\pgfpathclose
\pgfusepathqclip
\begin{pgfscope}
\definecolor{fc}{rgb}{0.5000,0.5000,0.5000}
\pgfsetfillcolor{fc}
\pgfsetlinewidth{0.8248bp}
\definecolor{sc}{rgb}{0.0000,0.0000,0.0000}
\pgfsetstrokecolor{sc}
\pgfsetmiterjoin
\pgfsetbuttcap
\pgfpathqmoveto{196.0000bp}{16.3333bp}
\pgfpathqlineto{196.0000bp}{36.7500bp}
\pgfpathqlineto{134.7500bp}{36.7500bp}
\pgfpathqlineto{134.7500bp}{16.3333bp}
\pgfpathqlineto{196.0000bp}{16.3333bp}
\pgfpathclose
\pgfusepathqfillstroke
\end{pgfscope}
\end{pgfscope}
\begin{pgfscope}
\definecolor{fc}{rgb}{0.5000,0.5000,0.5000}
\pgfsetfillcolor{fc}
\pgfsetlinewidth{0.8248bp}
\definecolor{sc}{rgb}{0.0000,0.0000,0.0000}
\pgfsetstrokecolor{sc}
\pgfsetmiterjoin
\pgfsetbuttcap
\pgfpathqmoveto{114.3333bp}{16.3333bp}
\pgfpathqlineto{114.3333bp}{36.7500bp}
\pgfpathqlineto{104.1250bp}{36.7500bp}
\pgfpathqlineto{104.1250bp}{16.3333bp}
\pgfpathqlineto{114.3333bp}{16.3333bp}
\pgfpathclose
\pgfusepathqfillstroke
\end{pgfscope}
\end{pgfscope}
\begin{pgfscope}
\pgfpathqmoveto{165.3750bp}{0.0000bp}
\pgfpathqlineto{165.3750bp}{61.2500bp}
\pgfpathqlineto{22.4583bp}{61.2500bp}
\pgfpathqlineto{22.4583bp}{0.0000bp}
\pgfpathqlineto{165.3750bp}{0.0000bp}
\pgfpathclose
\pgfusepathqclip
\begin{pgfscope}
\definecolor{fc}{rgb}{0.5000,0.5000,0.5000}
\pgfsetfillcolor{fc}
\pgfsetlinewidth{0.8248bp}
\definecolor{sc}{rgb}{0.0000,0.0000,0.0000}
\pgfsetstrokecolor{sc}
\pgfsetmiterjoin
\pgfsetbuttcap
\pgfpathqmoveto{83.7083bp}{16.3333bp}
\pgfpathqlineto{83.7083bp}{36.7500bp}
\pgfpathqlineto{73.5000bp}{36.7500bp}
\pgfpathqlineto{73.5000bp}{16.3333bp}
\pgfpathqlineto{83.7083bp}{16.3333bp}
\pgfpathclose
\pgfusepathqfillstroke
\end{pgfscope}
\end{pgfscope}
\begin{pgfscope}
\definecolor{fc}{rgb}{0.5000,0.5000,0.5000}
\pgfsetfillcolor{fc}
\pgfsetlinewidth{0.8248bp}
\definecolor{sc}{rgb}{0.0000,0.0000,0.0000}
\pgfsetstrokecolor{sc}
\pgfsetmiterjoin
\pgfsetbuttcap
\pgfpathqmoveto{53.0833bp}{16.3333bp}
\pgfpathqlineto{53.0833bp}{36.7500bp}
\pgfpathqlineto{12.2500bp}{36.7500bp}
\pgfpathqlineto{12.2500bp}{16.3333bp}
\pgfpathqlineto{53.0833bp}{16.3333bp}
\pgfpathclose
\pgfusepathqfillstroke
\end{pgfscope}
\end{pgfscope}
\end{pgfscope}
\begin{pgfscope}
\pgfpathqmoveto{165.3750bp}{0.0000bp}
\pgfpathqlineto{165.3750bp}{61.2500bp}
\pgfpathqlineto{22.4583bp}{61.2500bp}
\pgfpathqlineto{22.4583bp}{0.0000bp}
\pgfpathqlineto{165.3750bp}{0.0000bp}
\pgfpathclose
\pgfusepathqclip
\begin{pgfscope}
\definecolor{fc}{rgb}{1.0000,1.0000,1.0000}
\pgfsetfillcolor{fc}
\pgfsetfillopacity{0.7000}
\pgfsetlinewidth{2.0000bp}
\definecolor{sc}{rgb}{0.0000,0.0000,0.0000}
\pgfsetstrokecolor{sc}
\pgfsetmiterjoin
\pgfsetbuttcap
\pgfpathqmoveto{267.4583bp}{-167.4167bp}
\pgfpathqlineto{267.4583bp}{240.9167bp}
\pgfpathqlineto{-140.8750bp}{240.9167bp}
\pgfpathqlineto{-140.8750bp}{-167.4167bp}
\pgfpathqlineto{267.4583bp}{-167.4167bp}
\pgfpathclose
\pgfpathqmoveto{79.6250bp}{36.7500bp}
\pgfpathqcurveto{79.6250bp}{27.7293bp}{72.3123bp}{20.4167bp}{63.2917bp}{20.4167bp}
\pgfpathqcurveto{54.2710bp}{20.4167bp}{46.9583bp}{27.7293bp}{46.9583bp}{36.7500bp}
\pgfpathqcurveto{46.9583bp}{45.7707bp}{54.2710bp}{53.0833bp}{63.2917bp}{53.0833bp}
\pgfpathqcurveto{72.3123bp}{53.0833bp}{79.6250bp}{45.7707bp}{79.6250bp}{36.7500bp}
\pgfpathclose
\pgfusepathqfillstroke
\end{pgfscope}
\end{pgfscope}
\end{pgfscope}
\begin{pgfscope}
\definecolor{fc}{rgb}{0.0000,0.0000,0.0000}
\pgfsetfillcolor{fc}
\pgftransformcm{1.0000}{0.0000}{0.0000}{1.0000}{\pgfqpoint{17.3542bp}{30.6250bp}}
\pgftransformscale{1.0000}
\pgftext[right]{(c)}
\end{pgfscope}
\begin{pgfscope}
\definecolor{fc}{rgb}{0.2000,0.2000,0.2000}
\pgfsetfillcolor{fc}
\pgfpathqmoveto{142.4062bp}{90.8542bp}
\pgfpathqlineto{142.4062bp}{111.2708bp}
\pgfpathqcurveto{142.4062bp}{113.5260bp}{140.5781bp}{115.3542bp}{138.3229bp}{115.3542bp}
\pgfpathqlineto{108.2083bp}{115.3542bp}
\pgfpathqcurveto{105.9532bp}{115.3542bp}{104.1250bp}{113.5260bp}{104.1250bp}{111.2708bp}
\pgfpathqlineto{104.1250bp}{90.8542bp}
\pgfpathqcurveto{104.1250bp}{88.5990bp}{105.9532bp}{86.7708bp}{108.2083bp}{86.7708bp}
\pgfpathqlineto{138.3229bp}{86.7708bp}
\pgfpathqcurveto{140.5781bp}{86.7708bp}{142.4062bp}{88.5990bp}{142.4062bp}{90.8542bp}
\pgfpathclose
\pgfusepathqfill
\end{pgfscope}
\begin{pgfscope}
\definecolor{fc}{rgb}{1.0000,1.0000,0.0000}
\pgfsetfillcolor{fc}
\pgfsetlinewidth{0.8248bp}
\definecolor{sc}{rgb}{0.0000,0.0000,0.0000}
\pgfsetstrokecolor{sc}
\pgfsetmiterjoin
\pgfsetbuttcap
\pgfpathqmoveto{124.5417bp}{81.6667bp}
\pgfpathqlineto{124.5417bp}{111.2708bp}
\pgfpathqlineto{114.3333bp}{111.2708bp}
\pgfpathqlineto{114.3333bp}{81.6667bp}
\pgfpathqlineto{124.5417bp}{81.6667bp}
\pgfpathclose
\pgfusepathqfillstroke
\end{pgfscope}
\begin{pgfscope}
\definecolor{fc}{rgb}{0.2000,0.2000,0.2000}
\pgfsetfillcolor{fc}
\pgfpathqmoveto{70.9479bp}{90.8542bp}
\pgfpathqlineto{70.9479bp}{111.2708bp}
\pgfpathqcurveto{70.9479bp}{113.5260bp}{69.1197bp}{115.3542bp}{66.8646bp}{115.3542bp}
\pgfpathqlineto{36.7500bp}{115.3542bp}
\pgfpathqcurveto{34.4948bp}{115.3542bp}{32.6667bp}{113.5260bp}{32.6667bp}{111.2708bp}
\pgfpathqlineto{32.6667bp}{90.8542bp}
\pgfpathqcurveto{32.6667bp}{88.5990bp}{34.4948bp}{86.7708bp}{36.7500bp}{86.7708bp}
\pgfpathqlineto{66.8646bp}{86.7708bp}
\pgfpathqcurveto{69.1197bp}{86.7708bp}{70.9479bp}{88.5990bp}{70.9479bp}{90.8542bp}
\pgfpathclose
\pgfusepathqfill
\end{pgfscope}
\begin{pgfscope}
\definecolor{fc}{rgb}{1.0000,1.0000,0.0000}
\pgfsetfillcolor{fc}
\pgfsetlinewidth{0.8248bp}
\definecolor{sc}{rgb}{0.0000,0.0000,0.0000}
\pgfsetstrokecolor{sc}
\pgfsetmiterjoin
\pgfsetbuttcap
\pgfpathqmoveto{53.0833bp}{81.6667bp}
\pgfpathqlineto{53.0833bp}{111.2708bp}
\pgfpathqlineto{42.8750bp}{111.2708bp}
\pgfpathqlineto{42.8750bp}{81.6667bp}
\pgfpathqlineto{53.0833bp}{81.6667bp}
\pgfpathclose
\pgfusepathqfillstroke
\end{pgfscope}
\begin{pgfscope}
\definecolor{fc}{rgb}{0.6784,0.8471,0.9020}
\pgfsetfillcolor{fc}
\pgfsetlinewidth{0.8248bp}
\definecolor{sc}{rgb}{0.0000,0.0000,0.0000}
\pgfsetstrokecolor{sc}
\pgfsetmiterjoin
\pgfsetbuttcap
\pgfpathqmoveto{165.3750bp}{61.2500bp}
\pgfpathqlineto{165.3750bp}{81.6667bp}
\pgfpathqlineto{22.4583bp}{81.6667bp}
\pgfpathqlineto{22.4583bp}{61.2500bp}
\pgfpathqlineto{165.3750bp}{61.2500bp}
\pgfpathclose
\pgfusepathqfillstroke
\end{pgfscope}
\begin{pgfscope}
\definecolor{fc}{rgb}{0.8667,0.6275,0.8667}
\pgfsetfillcolor{fc}
\pgfsetlinewidth{0.8248bp}
\definecolor{sc}{rgb}{0.0000,0.0000,0.0000}
\pgfsetstrokecolor{sc}
\pgfsetmiterjoin
\pgfsetbuttcap
\pgfpathqmoveto{165.3750bp}{81.6667bp}
\pgfpathqlineto{165.3750bp}{91.8750bp}
\pgfpathqlineto{124.5417bp}{91.8750bp}
\pgfpathqlineto{124.5417bp}{81.6667bp}
\pgfpathqlineto{165.3750bp}{81.6667bp}
\pgfpathclose
\pgfusepathqfillstroke
\end{pgfscope}
\begin{pgfscope}
\definecolor{fc}{rgb}{0.8667,0.6275,0.8667}
\pgfsetfillcolor{fc}
\pgfsetlinewidth{0.8248bp}
\definecolor{sc}{rgb}{0.0000,0.0000,0.0000}
\pgfsetstrokecolor{sc}
\pgfsetmiterjoin
\pgfsetbuttcap
\pgfpathqmoveto{114.3333bp}{81.6667bp}
\pgfpathqlineto{114.3333bp}{91.8750bp}
\pgfpathqlineto{53.0833bp}{91.8750bp}
\pgfpathqlineto{53.0833bp}{81.6667bp}
\pgfpathqlineto{114.3333bp}{81.6667bp}
\pgfpathclose
\pgfusepathqfillstroke
\end{pgfscope}
\begin{pgfscope}
\definecolor{fc}{rgb}{0.8667,0.6275,0.8667}
\pgfsetfillcolor{fc}
\pgfsetlinewidth{0.8248bp}
\definecolor{sc}{rgb}{0.0000,0.0000,0.0000}
\pgfsetstrokecolor{sc}
\pgfsetmiterjoin
\pgfsetbuttcap
\pgfpathqmoveto{42.8750bp}{81.6667bp}
\pgfpathqlineto{42.8750bp}{91.8750bp}
\pgfpathqlineto{22.4583bp}{91.8750bp}
\pgfpathqlineto{22.4583bp}{81.6667bp}
\pgfpathqlineto{42.8750bp}{81.6667bp}
\pgfpathclose
\pgfusepathqfillstroke
\end{pgfscope}
\begin{pgfscope}
\definecolor{fc}{rgb}{0.1176,0.5647,1.0000}
\pgfsetfillcolor{fc}
\pgfsetlinewidth{0.8248bp}
\definecolor{sc}{rgb}{0.0000,0.0000,0.0000}
\pgfsetstrokecolor{sc}
\pgfsetmiterjoin
\pgfsetbuttcap
\pgfpathqmoveto{165.3750bp}{91.8750bp}
\pgfpathqlineto{165.3750bp}{102.0833bp}
\pgfpathqlineto{155.1667bp}{102.0833bp}
\pgfpathqlineto{155.1667bp}{91.8750bp}
\pgfpathqlineto{165.3750bp}{91.8750bp}
\pgfpathclose
\pgfusepathqfillstroke
\end{pgfscope}
\begin{pgfscope}
\definecolor{fc}{rgb}{0.1176,0.5647,1.0000}
\pgfsetfillcolor{fc}
\pgfsetlinewidth{0.8248bp}
\definecolor{sc}{rgb}{0.0000,0.0000,0.0000}
\pgfsetstrokecolor{sc}
\pgfsetmiterjoin
\pgfsetbuttcap
\pgfpathqmoveto{144.9583bp}{91.8750bp}
\pgfpathqlineto{144.9583bp}{102.0833bp}
\pgfpathqlineto{139.8542bp}{102.0833bp}
\pgfpathqlineto{139.8542bp}{91.8750bp}
\pgfpathqlineto{144.9583bp}{91.8750bp}
\pgfpathclose
\pgfusepathqfillstroke
\end{pgfscope}
\begin{pgfscope}
\definecolor{fc}{rgb}{0.1176,0.5647,1.0000}
\pgfsetfillcolor{fc}
\pgfsetlinewidth{0.8248bp}
\definecolor{sc}{rgb}{0.0000,0.0000,0.0000}
\pgfsetstrokecolor{sc}
\pgfsetmiterjoin
\pgfsetbuttcap
\pgfpathqmoveto{129.6458bp}{91.8750bp}
\pgfpathqlineto{129.6458bp}{102.0833bp}
\pgfpathqlineto{124.5417bp}{102.0833bp}
\pgfpathqlineto{124.5417bp}{91.8750bp}
\pgfpathqlineto{129.6458bp}{91.8750bp}
\pgfpathclose
\pgfusepathqfillstroke
\end{pgfscope}
\begin{pgfscope}
\definecolor{fc}{rgb}{0.1176,0.5647,1.0000}
\pgfsetfillcolor{fc}
\pgfsetlinewidth{0.8248bp}
\definecolor{sc}{rgb}{0.0000,0.0000,0.0000}
\pgfsetstrokecolor{sc}
\pgfsetmiterjoin
\pgfsetbuttcap
\pgfpathqmoveto{114.3333bp}{91.8750bp}
\pgfpathqlineto{114.3333bp}{102.0833bp}
\pgfpathqlineto{83.7083bp}{102.0833bp}
\pgfpathqlineto{83.7083bp}{91.8750bp}
\pgfpathqlineto{114.3333bp}{91.8750bp}
\pgfpathclose
\pgfusepathqfillstroke
\end{pgfscope}
\begin{pgfscope}
\definecolor{fc}{rgb}{0.1176,0.5647,1.0000}
\pgfsetfillcolor{fc}
\pgfsetlinewidth{0.8248bp}
\definecolor{sc}{rgb}{0.0000,0.0000,0.0000}
\pgfsetstrokecolor{sc}
\pgfsetmiterjoin
\pgfsetbuttcap
\pgfpathqmoveto{73.5000bp}{91.8750bp}
\pgfpathqlineto{73.5000bp}{102.0833bp}
\pgfpathqlineto{68.3958bp}{102.0833bp}
\pgfpathqlineto{68.3958bp}{91.8750bp}
\pgfpathqlineto{73.5000bp}{91.8750bp}
\pgfpathclose
\pgfusepathqfillstroke
\end{pgfscope}
\begin{pgfscope}
\definecolor{fc}{rgb}{0.1176,0.5647,1.0000}
\pgfsetfillcolor{fc}
\pgfsetlinewidth{0.8248bp}
\definecolor{sc}{rgb}{0.0000,0.0000,0.0000}
\pgfsetstrokecolor{sc}
\pgfsetmiterjoin
\pgfsetbuttcap
\pgfpathqmoveto{58.1875bp}{91.8750bp}
\pgfpathqlineto{58.1875bp}{102.0833bp}
\pgfpathqlineto{53.0833bp}{102.0833bp}
\pgfpathqlineto{53.0833bp}{91.8750bp}
\pgfpathqlineto{58.1875bp}{91.8750bp}
\pgfpathclose
\pgfusepathqfillstroke
\end{pgfscope}
\begin{pgfscope}
\definecolor{fc}{rgb}{0.1176,0.5647,1.0000}
\pgfsetfillcolor{fc}
\pgfsetlinewidth{0.8248bp}
\definecolor{sc}{rgb}{0.0000,0.0000,0.0000}
\pgfsetstrokecolor{sc}
\pgfsetmiterjoin
\pgfsetbuttcap
\pgfpathqmoveto{42.8750bp}{91.8750bp}
\pgfpathqlineto{42.8750bp}{102.0833bp}
\pgfpathqlineto{22.4583bp}{102.0833bp}
\pgfpathqlineto{22.4583bp}{91.8750bp}
\pgfpathqlineto{42.8750bp}{91.8750bp}
\pgfpathclose
\pgfusepathqfillstroke
\end{pgfscope}
\begin{pgfscope}
\definecolor{fc}{rgb}{0.5000,0.5000,0.5000}
\pgfsetfillcolor{fc}
\pgfsetlinewidth{0.8248bp}
\definecolor{sc}{rgb}{0.0000,0.0000,0.0000}
\pgfsetstrokecolor{sc}
\pgfsetmiterjoin
\pgfsetbuttcap
\pgfpathqmoveto{165.3750bp}{102.0833bp}
\pgfpathqlineto{165.3750bp}{112.2917bp}
\pgfpathqlineto{155.1667bp}{112.2917bp}
\pgfpathqlineto{155.1667bp}{102.0833bp}
\pgfpathqlineto{165.3750bp}{102.0833bp}
\pgfpathclose
\pgfusepathqfillstroke
\end{pgfscope}
\begin{pgfscope}
\definecolor{fc}{rgb}{0.5000,0.5000,0.5000}
\pgfsetfillcolor{fc}
\pgfsetlinewidth{0.8248bp}
\definecolor{sc}{rgb}{0.0000,0.0000,0.0000}
\pgfsetstrokecolor{sc}
\pgfsetmiterjoin
\pgfsetbuttcap
\pgfpathqmoveto{144.9583bp}{102.0833bp}
\pgfpathqlineto{144.9583bp}{112.2917bp}
\pgfpathqlineto{139.8542bp}{112.2917bp}
\pgfpathqlineto{139.8542bp}{102.0833bp}
\pgfpathqlineto{144.9583bp}{102.0833bp}
\pgfpathclose
\pgfusepathqfillstroke
\end{pgfscope}
\begin{pgfscope}
\definecolor{fc}{rgb}{0.5000,0.5000,0.5000}
\pgfsetfillcolor{fc}
\pgfsetlinewidth{0.8248bp}
\definecolor{sc}{rgb}{0.0000,0.0000,0.0000}
\pgfsetstrokecolor{sc}
\pgfsetmiterjoin
\pgfsetbuttcap
\pgfpathqmoveto{129.6458bp}{102.0833bp}
\pgfpathqlineto{129.6458bp}{112.2917bp}
\pgfpathqlineto{124.5417bp}{112.2917bp}
\pgfpathqlineto{124.5417bp}{102.0833bp}
\pgfpathqlineto{129.6458bp}{102.0833bp}
\pgfpathclose
\pgfusepathqfillstroke
\end{pgfscope}
\begin{pgfscope}
\definecolor{fc}{rgb}{0.5000,0.5000,0.5000}
\pgfsetfillcolor{fc}
\pgfsetlinewidth{0.8248bp}
\definecolor{sc}{rgb}{0.0000,0.0000,0.0000}
\pgfsetstrokecolor{sc}
\pgfsetmiterjoin
\pgfsetbuttcap
\pgfpathqmoveto{114.3333bp}{102.0833bp}
\pgfpathqlineto{114.3333bp}{112.2917bp}
\pgfpathqlineto{83.7083bp}{112.2917bp}
\pgfpathqlineto{83.7083bp}{102.0833bp}
\pgfpathqlineto{114.3333bp}{102.0833bp}
\pgfpathclose
\pgfusepathqfillstroke
\end{pgfscope}
\begin{pgfscope}
\definecolor{fc}{rgb}{0.5000,0.5000,0.5000}
\pgfsetfillcolor{fc}
\pgfsetlinewidth{0.8248bp}
\definecolor{sc}{rgb}{0.0000,0.0000,0.0000}
\pgfsetstrokecolor{sc}
\pgfsetmiterjoin
\pgfsetbuttcap
\pgfpathqmoveto{73.5000bp}{102.0833bp}
\pgfpathqlineto{73.5000bp}{112.2917bp}
\pgfpathqlineto{68.3958bp}{112.2917bp}
\pgfpathqlineto{68.3958bp}{102.0833bp}
\pgfpathqlineto{73.5000bp}{102.0833bp}
\pgfpathclose
\pgfusepathqfillstroke
\end{pgfscope}
\begin{pgfscope}
\definecolor{fc}{rgb}{0.5000,0.5000,0.5000}
\pgfsetfillcolor{fc}
\pgfsetlinewidth{0.8248bp}
\definecolor{sc}{rgb}{0.0000,0.0000,0.0000}
\pgfsetstrokecolor{sc}
\pgfsetmiterjoin
\pgfsetbuttcap
\pgfpathqmoveto{58.1875bp}{102.0833bp}
\pgfpathqlineto{58.1875bp}{112.2917bp}
\pgfpathqlineto{53.0833bp}{112.2917bp}
\pgfpathqlineto{53.0833bp}{102.0833bp}
\pgfpathqlineto{58.1875bp}{102.0833bp}
\pgfpathclose
\pgfusepathqfillstroke
\end{pgfscope}
\begin{pgfscope}
\definecolor{fc}{rgb}{0.5000,0.5000,0.5000}
\pgfsetfillcolor{fc}
\pgfsetlinewidth{0.8248bp}
\definecolor{sc}{rgb}{0.0000,0.0000,0.0000}
\pgfsetstrokecolor{sc}
\pgfsetmiterjoin
\pgfsetbuttcap
\pgfpathqmoveto{42.8750bp}{102.0833bp}
\pgfpathqlineto{42.8750bp}{112.2917bp}
\pgfpathqlineto{22.4583bp}{112.2917bp}
\pgfpathqlineto{22.4583bp}{102.0833bp}
\pgfpathqlineto{42.8750bp}{102.0833bp}
\pgfpathclose
\pgfusepathqfillstroke
\end{pgfscope}
\begin{pgfscope}
\definecolor{fc}{rgb}{0.0000,0.0000,0.0000}
\pgfsetfillcolor{fc}
\pgftransformcm{1.0000}{0.0000}{0.0000}{1.0000}{\pgfqpoint{172.5208bp}{107.1875bp}}
\pgftransformscale{1.0000}
\pgftext[left]{inkjet conductor}
\end{pgfscope}
\begin{pgfscope}
\pgfsetlinewidth{0.8248bp}
\definecolor{sc}{rgb}{0.0000,0.0000,0.0000}
\pgfsetstrokecolor{sc}
\pgfsetmiterjoin
\pgfsetbuttcap
\pgfpathqmoveto{134.7500bp}{107.1875bp}
\pgfpathqlineto{170.4792bp}{107.1875bp}
\pgfusepathqstroke
\end{pgfscope}
\begin{pgfscope}
\definecolor{fc}{rgb}{0.0000,0.0000,0.0000}
\pgfsetfillcolor{fc}
\pgftransformcm{1.0000}{0.0000}{0.0000}{1.0000}{\pgfqpoint{172.5208bp}{86.7708bp}}
\pgftransformscale{1.0000}
\pgftext[left]{inkjet insulator}
\end{pgfscope}
\begin{pgfscope}
\pgfsetlinewidth{0.8248bp}
\definecolor{sc}{rgb}{0.0000,0.0000,0.0000}
\pgfsetstrokecolor{sc}
\pgfsetmiterjoin
\pgfsetbuttcap
\pgfpathqmoveto{119.4375bp}{86.7708bp}
\pgfpathqlineto{170.4792bp}{86.7708bp}
\pgfusepathqstroke
\end{pgfscope}
\begin{pgfscope}
\definecolor{fc}{rgb}{0.0000,0.0000,0.0000}
\pgfsetfillcolor{fc}
\pgftransformcm{1.0000}{0.0000}{0.0000}{1.0000}{\pgfqpoint{17.3542bp}{88.3021bp}}
\pgftransformscale{1.0000}
\pgftext[right]{(b)}
\end{pgfscope}
\begin{pgfscope}
\definecolor{fc}{rgb}{0.6784,0.8471,0.9020}
\pgfsetfillcolor{fc}
\pgfsetlinewidth{0.8248bp}
\definecolor{sc}{rgb}{0.0000,0.0000,0.0000}
\pgfsetstrokecolor{sc}
\pgfsetmiterjoin
\pgfsetbuttcap
\pgfpathqmoveto{165.3750bp}{122.5000bp}
\pgfpathqlineto{165.3750bp}{142.9167bp}
\pgfpathqlineto{22.4583bp}{142.9167bp}
\pgfpathqlineto{22.4583bp}{122.5000bp}
\pgfpathqlineto{165.3750bp}{122.5000bp}
\pgfpathclose
\pgfusepathqfillstroke
\end{pgfscope}
\begin{pgfscope}
\definecolor{fc}{rgb}{0.0000,0.0000,0.0000}
\pgfsetfillcolor{fc}
\pgftransformcm{1.0000}{0.0000}{0.0000}{1.0000}{\pgfqpoint{172.5208bp}{132.7083bp}}
\pgftransformscale{1.0000}
\pgftext[left]{glass}
\end{pgfscope}
\begin{pgfscope}
\pgfsetlinewidth{0.8248bp}
\definecolor{sc}{rgb}{0.0000,0.0000,0.0000}
\pgfsetstrokecolor{sc}
\pgfsetmiterjoin
\pgfsetbuttcap
\pgfpathqmoveto{160.2708bp}{132.7083bp}
\pgfpathqlineto{170.4792bp}{132.7083bp}
\pgfusepathqstroke
\end{pgfscope}
\begin{pgfscope}
\definecolor{fc}{rgb}{0.8667,0.6275,0.8667}
\pgfsetfillcolor{fc}
\pgfsetlinewidth{0.8248bp}
\definecolor{sc}{rgb}{0.0000,0.0000,0.0000}
\pgfsetstrokecolor{sc}
\pgfsetmiterjoin
\pgfsetbuttcap
\pgfpathqmoveto{165.3750bp}{142.9167bp}
\pgfpathqlineto{165.3750bp}{153.1250bp}
\pgfpathqlineto{22.4583bp}{153.1250bp}
\pgfpathqlineto{22.4583bp}{142.9167bp}
\pgfpathqlineto{165.3750bp}{142.9167bp}
\pgfpathclose
\pgfusepathqfillstroke
\end{pgfscope}
\begin{pgfscope}
\definecolor{fc}{rgb}{0.0000,0.0000,0.0000}
\pgfsetfillcolor{fc}
\pgftransformcm{1.0000}{0.0000}{0.0000}{1.0000}{\pgfqpoint{172.5208bp}{148.0208bp}}
\pgftransformscale{1.0000}
\pgftext[left]{TCO}
\end{pgfscope}
\begin{pgfscope}
\pgfsetlinewidth{0.8248bp}
\definecolor{sc}{rgb}{0.0000,0.0000,0.0000}
\pgfsetstrokecolor{sc}
\pgfsetmiterjoin
\pgfsetbuttcap
\pgfpathqmoveto{160.2708bp}{148.0208bp}
\pgfpathqlineto{170.4792bp}{148.0208bp}
\pgfusepathqstroke
\end{pgfscope}
\begin{pgfscope}
\definecolor{fc}{rgb}{0.1176,0.5647,1.0000}
\pgfsetfillcolor{fc}
\pgfsetlinewidth{0.8248bp}
\definecolor{sc}{rgb}{0.0000,0.0000,0.0000}
\pgfsetstrokecolor{sc}
\pgfsetmiterjoin
\pgfsetbuttcap
\pgfpathqmoveto{165.3750bp}{153.1250bp}
\pgfpathqlineto{165.3750bp}{163.3333bp}
\pgfpathqlineto{22.4583bp}{163.3333bp}
\pgfpathqlineto{22.4583bp}{153.1250bp}
\pgfpathqlineto{165.3750bp}{153.1250bp}
\pgfpathclose
\pgfusepathqfillstroke
\end{pgfscope}
\begin{pgfscope}
\definecolor{fc}{rgb}{0.0000,0.0000,0.0000}
\pgfsetfillcolor{fc}
\pgftransformcm{1.0000}{0.0000}{0.0000}{1.0000}{\pgfqpoint{172.5208bp}{158.2292bp}}
\pgftransformscale{1.0000}
\pgftext[left]{semiconductor}
\end{pgfscope}
\begin{pgfscope}
\pgfsetlinewidth{0.8248bp}
\definecolor{sc}{rgb}{0.0000,0.0000,0.0000}
\pgfsetstrokecolor{sc}
\pgfsetmiterjoin
\pgfsetbuttcap
\pgfpathqmoveto{160.2708bp}{158.2292bp}
\pgfpathqlineto{170.4792bp}{158.2292bp}
\pgfusepathqstroke
\end{pgfscope}
\begin{pgfscope}
\definecolor{fc}{rgb}{0.5000,0.5000,0.5000}
\pgfsetfillcolor{fc}
\pgfsetlinewidth{0.8248bp}
\definecolor{sc}{rgb}{0.0000,0.0000,0.0000}
\pgfsetstrokecolor{sc}
\pgfsetmiterjoin
\pgfsetbuttcap
\pgfpathqmoveto{165.3750bp}{163.3333bp}
\pgfpathqlineto{165.3750bp}{173.5417bp}
\pgfpathqlineto{22.4583bp}{173.5417bp}
\pgfpathqlineto{22.4583bp}{163.3333bp}
\pgfpathqlineto{165.3750bp}{163.3333bp}
\pgfpathclose
\pgfusepathqfillstroke
\end{pgfscope}
\begin{pgfscope}
\definecolor{fc}{rgb}{0.0000,0.0000,0.0000}
\pgfsetfillcolor{fc}
\pgftransformcm{1.0000}{0.0000}{0.0000}{1.0000}{\pgfqpoint{172.5208bp}{168.4375bp}}
\pgftransformscale{1.0000}
\pgftext[left]{metal}
\end{pgfscope}
\begin{pgfscope}
\pgfsetlinewidth{0.8248bp}
\definecolor{sc}{rgb}{0.0000,0.0000,0.0000}
\pgfsetstrokecolor{sc}
\pgfsetmiterjoin
\pgfsetbuttcap
\pgfpathqmoveto{160.2708bp}{168.4375bp}
\pgfpathqlineto{170.4792bp}{168.4375bp}
\pgfusepathqstroke
\end{pgfscope}
\begin{pgfscope}
\definecolor{fc}{rgb}{0.0000,0.0000,0.0000}
\pgfsetfillcolor{fc}
\pgftransformcm{1.0000}{0.0000}{0.0000}{1.0000}{\pgfqpoint{17.3542bp}{148.0208bp}}
\pgftransformscale{1.0000}
\pgftext[right]{(a)}
\end{pgfscope}
\end{pgfpicture}

\caption{%
  Illustration of the solar cell manufacturing process developed in
  Ref.~\cite{walls}, which uses inkjet printing. In the initial stage
  (a), three layers are deposited in sequence onto a glass substrate.
  The first is a transparent conducting oxide (TCO) layer, then the
  semiconductor and finally a metallic layer. Then, trenches are made by
  depth selective laser scribes. Following this, the insulating polymer
  and conductive inks are deposited, as shown in (b). When the process
  is complete, the conductive ink should form a conducting connection
  between the TCO at the bottom of one cell and the metal on the top of
  the neighbouring cell, bridging the insulating polymer. In (c) we
  display a zoom of the conducting connection that we model here.
}
\label{fig:solar}
\end{figure}

The surface onto which the nanoparticle ink is printed consists of two
materials: (i) a metal conducting surface that is either side of (ii) a
strip of a polymer insulating material. The metallic part of the surface
is hydrophilic and the polymeric part is hydrophobic. When the liquid is
deposited onto these two materials side-by-side, there is a tendency for
the liquid to dewet from the surface of the insulator and move onto
the metal, since this reduces the energy of the system. In the solar
cell manufacturing process~\cite{walls}, this insulating polymer strip
is created by inkjet printing into a trench created on the surface by
laser ablation, at a previous stage --- see Fig.~\ref{fig:solar}.

If the nanoparticle suspension deposited perpendicular to the polymer strip is to
dry to form an electrical connection, it is crucial that the ink does
not dewet from the hydrophobic surface. The aim of the present work is
to understand when this dewetting occurs and also to determine if there
are processes that can be done during manufacturing 
to prevent dewetting.

The specific example considered is a particular case of a more general
class of problem: that of modelling the evaporation of a nanoparticle
suspension from a heterogeneous surface. The deposition and drying of
the ink involves processes that occur over a huge range of time and
length scales. The procedure can be roughly split into two parts:
(i) the process of the ink being sprayed from the print head and
arriving at the surface and (ii) the behaviour of the ink as it dries,
once it is on the surface. In our work, we focus solely on stage (ii),
in which there are still processes that occur over a great range of
length and time scales.

The nanoparticles move throughout the liquid with a diffusive dynamics,
where changes occur on a time scale much larger than the time scale for
rearrangements of the solvent molecules. Drop shape changes occur on a
time scale that is very much larger than the molecular time scale and
also the nanoparticle diffusive time scale. There are also several
disparate length scales, ranging from the solvent molecular diameter
scale, to the size of the nanoparticles, the scale of any surface
structures and then largest of all, the liquid drop size. Because of
this, modelling such a multi-scale system has many challenges.
Mesoscopic thin-film partial differential equation based models can be
used~\cite{oron1997, kalliadasis2007, thiele2009, dietrich, frastia2011,
frastia2012} but relating properties of the microscopic interactions
between the particles and the structures they form in the liquid is not
straightforward, because this type of model describes the distribution
of the nanoparticles over the surface via a height-averaged
concentration profile. This does not allow a description of the
variations in the nanoparticles density distribution in the direction
perpendicular to the surface. A fully microscopic approach, such as
molecular dynamics (MD) does include every aspect of the motion of the
particles and can be used to describe small liquid drops on a surface
\cite{ingebrigtsen2007, lu2013, tretyakov2013, becker2014}. Generally MD
is computationally infeasible even for moderate system sizes due to the
long time scales over which the evaporative drying occurs. Similarly,
classical density functional theory (DFT)~\cite{Evans79, Evans92,
hansen} and dynamical DFT~\cite{MaTa99, ArEv04, archer2006, archer2009}
can describe in great detail the density profile of the liquid at the
interface and the structure down to the scale of individual
particles~\cite{Evans92, hansen, nold2014, hughes2015, hughes2017}
but the level of detail makes this also computationally very expensive.

We require a coarse-grained model to describe the fluid dynamical
processes of interest here but not to the degree of coarse-graining as
is present in the thin-film equation based models. Thus, we develop a
lattice model for the system using { Monte Carlo (MC) to capture
the non-equilibrium} dynamics and model the system time evolution as a series of discrete
events. We model the nanoparticles individually, incorporating in the
model their diffusion through the liquid over time thus enabling a
description of the structures they may form on the surface. However,
instead of modelling every solvent molecule individually, we group them
together and statistically model them by a single, larger effective
`particle' of the same size as the nanoparticles, also residing on a
lattice. MC models of this type have been used before, initially by
treating the system effectively in two dimensions~\cite{rabani2003,
pauliac2008, vancea08, stannard2011dewetting}. However, more recently
models that are fully three-dimensional have been used~\cite{sztrum2005,
kim2011, jung2014, crivoi2014, tewes2016comparing}. Our model here is of
this kind but differs from previous studies in the manner in which  we
describe the particle interactions, allowing for correct modelling of
the (hemispherical) liquid drop shape. Additionally, the effect of
surface roughness is incorporated.

How a liquid wets a surface is characterised by the {\em spreading parameter}
$s$ \cite{de2004capillarity}. It is defined as the difference in the
surface tensions between the liquid, gas and the substrate:
\begin{equation}
  s = \gamma_{\text{sg}} - (\gamma_{\text{sl}}+\gamma_{\text{lg}}).
\end{equation}
The first term, $\gamma_{\text{sg}}$, is the excess free energy per unit
area of the substrate when dry (i.e.~in contact with the gas phase),
referred to as the solid-gas interfacial tension. The second term is the
excess free energy per unit area of the substrate when it is wet by a
thick film of the liquid and is the sum of the solid-liquid interfacial
tension $\gamma_{\text{sl}}$ and the liquid-gas interfacial tension
$\gamma_{\text{lg}}$. When $s>0$ the liquid seeks to spread over the
surface. In contrast, when $s<0$ the liquid only partially wets the
substrate, forming a drop with contact angle $\theta$. Young's
equation~\cite{de2004capillarity} relates the contact angle $\theta$ to
the interfacial tensions
\begin{equation}
\gamma_{\text{lg}}\cos\theta=\gamma_{\text{sg}}-\gamma_{\text{sl}}.
\end{equation}
Therefore, the contact angle and spreading parameter are related by
$s = \gamma_{\text{lg}}(\cos\theta - 1)$.
From our simulation results we can calculate the contact angle and also
determine how this depends on the parameters in our model. Thus, to
model a particular experiment, we have to find the contact angle of the
solvent on the particular material(s) in the substrate (many are
available in the literature) and then we select our model parameters to
match the experiments.

The remainder of this paper is laid out as follows: In
Sec.~\ref{methodology} we describe our model and
the MC algorithm for the dynamics. This section also
presents results for the model when no
nanoparticles are present, to illustrate the wetting behaviour
of the pure solvent liquid on a uniform planar surface. We determine
the dependance of the contact angle on the model parameters,
to enable selecting values to match experiments.
In Sec.~\ref{sec:bulk} we briefly present the bulk solvent fluid phase diagram.
In Sec.~\ref{sec:droplets} we present results for droplets containing nanoparticles
evaporating from a smooth planar surface and also show how to include
the effect of surface roughness by changing the fluid dynamics in the
vicinity of the surface.
Sec.~\ref{nanoparticles} presents results for the drying of the nanoparticle
suspension from a heterogeneous surface, with emphasis on the drying
of liquid bridges spanning a hydrophobic patch. Finally, in Sec.~\ref{conclusion}
we make a few concluding remarks.

\section{Lattice model for the system}\label{methodology}

\begin{figure}[t]
\centering
\includegraphics{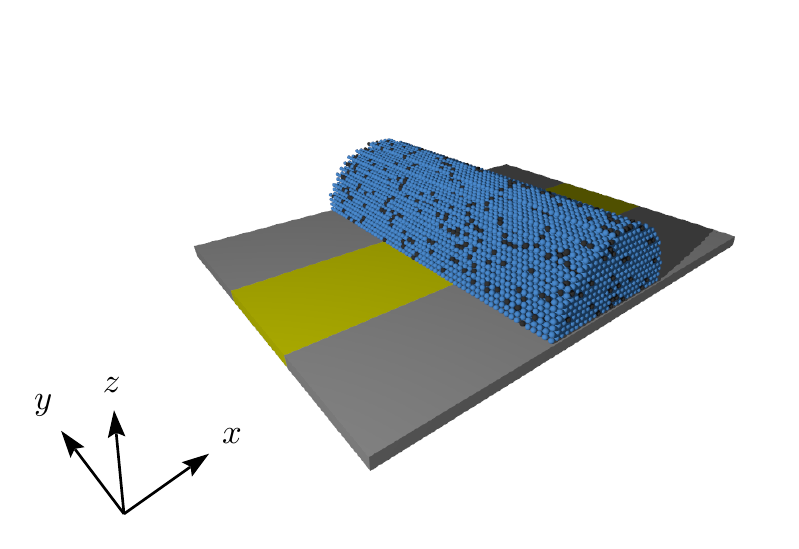}
\caption{%
  An example of a simulation starting condition. The liquid (blue) is
  initiated in a semi-circular strip connecting the conductive metal
  surfaces. Nanoparticles (black) are randomly dispersed throughout the liquid.
  The metal surface (in grey) can be elevated above the insulating
  portion in the middle (yellow), but in the case displayed here is not.
}
\label{fig:initial}
\end{figure}

The system is discretised onto a regular three dimensional grid with
lattice spacing $\sigma$ and with periodic boundary conditions in the
$x$ and $y$ coordinates. The surface of the substrate onto which the
nanoparticle suspension (ink)
is deposited is perpendicular to the $z$ direction. Ink cannot penetrate
the surface. At the top of the simulation box, at $z = L$, we apply
various different boundary conditions, discussed below. A typical
starting configuration is displayed in Fig.~\ref{fig:initial}. Each
lattice site above the surface can be in one of three states: (i) empty,
(ii) containing a nanoparticle or (iii) containing liquid. We refer to a
lattice site containing liquid as containing a liquid `particle', but it
should be borne in mind that this does not mean an individual solvent
molecule but rather many of them grouped together in a volume
$\sigma^3$. The lattice constant $\sigma$ is most easily envisaged as
being the diameter of the nanoparticles but this does not have to be so:
one can also consider $\sigma$ to be a larger coarse-graining length
scale, in which case when a lattice site is said to be `containing' a
nanoparticle, we mean `contains mostly nanoparticles'.

We define $n_{\i}$ and $l_{\i}$ to be the occupation numbers of lattice
site $\i$ for nanoparticles and liquid respectively, where
\begin{equation}
  \i = (i,j,k)
\end{equation}
is the discrete position vector (c.f.~Fig.~\ref{fig:initial}). If site
$\i$ is occupied by liquid, then $l_{\i}=1$, otherwise $l_{\i}=0$.
Similarly, if site $\i$ is occupied by a nanoparticle then $n_{\i}=1$
and $n_{\i} = 0$ if there is no nanoparticle. Liquid and a nanoparticle
cannot occupy the same site.

We model the total energy of the system $E$ by the following sum:
\begin{equation}
  \begin{aligned}
  E =& - \sum_{\i,\j} \left(
      \frac{\e_{nn}}{2}c_{\i\j} n_{\i}n_{\j}
    + \e_{nl}c_{\i\j} l_{\i}n_{\j}
    + \frac{\e_{ll}}{2}c_{\i\j} l_{\i}l_{\j}
  \right)\\
 &- \mu\sum_{\i} l_{\i}
  + \sum_{\i} V_{\i}^l l_{\i}
  + \sum_{\i} V_{\i}^n n_{\i}.
\end{aligned}
\label{eq:energy}
\end{equation}
The first term, a sum over pairs of lattice sites, is the contribution
from particle interactions. The overall strength of the interactions
between pairs of nanoparticles is determined by the parameter $\e_{nn}$,
between liquid and nanoparticles by $\e_{nl}$ and between pairs of
liquid particles by $\e_{ll}$. The precise value of the interaction
energy between pairs of particles at sites $\i$ and $\j$ is determined
by the dimensionless coefficient $c_{\i\j}$, which decreases in value as
the distance between the pair of particles increases. We use the
following values
\begin{equation}
  c_{\i\j} =
  \begin{dcases}
    1            & \j \in \{\text{NN } \i\}   \\
    \frac{3}{10} & \j \in \{\text{NNN } \i\}  \\
    \frac{1}{20} & \j \in \{\text{NNNN } \i\} \\
    0            & \text{otherwise}
  \end{dcases}
\label{eq:c_ij}
\end{equation}
where NN $\i$, NNN $\i$ and NNNN $\i$ stand for nearest neighbours, next
nearest neighbours and next-next nearest neighbours, respectively
{ Thus, we truncate all interactions for $|\i -\j|>\sqrt{3}\sigma$. The
influence on the wetting behaviour of truncating the range of the
interactions is discussed in Refs.~\cite{hughes2015,hughes2017}.}

The
choice of particular values in Eq.~\eqref{eq:c_ij} is important, as this
leads to liquid droplets on the surface having a hemispherical shape.
For example, if instead we set $c_{\i\j}=0$ for $\j \in$ NNN $\i$ and
$\j \in$ NNNN $\i$, (i.e.\ just nearest neighbour interactions) then the
system forms unrealistic rectangular shaped droplets, particularly at
low temperatures.
Thus, with the values in Eq.~\eqref{eq:c_ij} the dependence of the
gas-liquid surface tension on the orientation of the interface with
respect to the grid is minimised.
That one should select the particular values in Eq.~\eqref{eq:c_ij}
comes from noting that the sum over neighbours in Eq.~\eqref{eq:energy}
has the same form as a finite difference approximation for the Laplacian
\cite{robbins2011}. It can be shown that the values for $c_{\i\j}$ given
in Eq.~\eqref{eq:c_ij} minimise the errors from discretising the
Laplacian on the grid~\cite{isotropic}, dictating the choice in
Eq.~\eqref{eq:c_ij} \footnote{ In addition
to leading to hemispherical droplets, we expect the
particular values for $c_{\i\j}$ in Eq.~\eqref{eq:c_ij} to
result in the roughening transition
\cite{weeks1980roughening, abraham1980solvable} temperature
to be suppressed. Indeed, at all the temperatures for which we
have performed simulations, we have seen no evidence of a
roughening transition.}.

The second term in Eq.~(\ref{eq:energy}) is the contribution from
treating the liquid as being coupled to a reservoir, which is
the vapour above the surface. $\mu$ is the chemical potential of the
vapour. The value of $\mu$ determines the rate at which the liquid
evaporates from the surface. The last two terms of Eq.~\eqref{eq:energy}
are the contribution from the interaction with the surface, where
$V_{\i}^l$ and $V_\i^n$ are the external potentials due to the surface
exerted on the liquid and the nanoparticles, respectively. Assuming that
the surface is composed of particles interacting with the fluid with
interaction strength $\e_{wl}$ and a pair potential of the same form as
the pair potentials in Eq.~\eqref{eq:energy},
then for a flat structureless surface the potential takes the form
\begin{equation}
  V_{\i}^l =
  \begin{cases}
    \infty       & k < 1 \\
    -12\e_{wl}/5 & k = 1 \\
    0        & \text{otherwise,}
  \label{eq:v_il}
  \end{cases}
\end{equation}
where $k$ is the perpendicular distance from the surface. Similarly, if
the interaction strength with the nanoparticles is $\e_{wn}$, then the
external potential for the nanoparticles takes the same form
as~\eqref{eq:v_il} but the suffix $l$ is replaced by $n$.

The external potential in Eq.~\eqref{eq:v_il} is modified when the
surface varies in height or if the material changes. For example, to
model the situation illustrated in Fig.~\ref{fig:solar}, since the
polymer hydrophobic section of the surface is inkjet printed at an
earlier stage, its surface height can be controlled. We denote the step
in height from the polymer to the metal part of the surface, to be $h$.

In all that follows below, we non-dimensionalise and set $\e_{ll}$ to be
the unit of energy and the lattice spacing $\sigma$ to be our unit of
length. All other parameters are given in terms of these.


\subsection*{The Monte Carlo Algorithm}\label{monte-carlo}

We denote a particular state of the system as
$S_\alpha\equiv\{n_{\mathbf 1},n_{\mathbf 2},\ldots,l_{\mathbf
1},l_{\mathbf 2},\ldots\}$, i.e.~a particular set of values of the
occupation numbers, which we index with the label $\alpha$. We also
denote the probability of the system being in this state at time $t$ as
$P(S_\alpha,t)$. The time evolution of this probability is given by the
master equation
\begin{equation}
  \begin{aligned}
  P(S_\alpha,t+1)
    &= P(S_\alpha,t)
    -\sum_{\beta\neq\alpha} w_{\alpha\to\beta}P(S_\alpha,t) \\
    &\quad +\sum_{\beta\neq\alpha} w_{\beta\to\alpha}P(S_\beta,t)
  \end{aligned}
\end{equation}
where $w_{\alpha\to\beta}$ is the transition rate from state $S_\alpha$
to state $S_\beta$.

In equilibrium, where $P(S_\alpha,t+1)=P(S_\alpha,t)$, we have
\cite{landau2014guide}:
\begin{equation}\label{eq:ratio}
  \frac{w_{\alpha\to\beta}}{w_{\beta\to\alpha}}=e^{-\Delta E/k_{B}T}
\end{equation}
where $T$ is the temperature, $k_B$ is Boltzmann's constant and $\Delta
E=E(S_\beta)-E(S_\alpha)$, with the energy $E$ given in
Eq.~\eqref{eq:energy}. The following Metropolis Monte Carlo algorithm
satisfies this~\cite{landau2014guide}:
\begin{enumerate}
\item
  Pick a random particle.
\item
  Pick a random neighbouring site.
\item
  Calculate the change in energy, $\Delta E$, from swapping these
  particles using Eq.~\eqref{eq:energy}.
\item
  If $\Delta E < 0$, perform the swap. Otherwise, swap the particles
  with probability $e^{-\Delta E / k_B T}$.
\end{enumerate}

The assumption { here} is that even when the system is out of
equilibrium, the transition rates given by Eq.~\eqref{eq:ratio} still
hold, allowing us to use the algorithm to determine the non-equilibrium
dynamics of the { liquid}.

This algorithm is refined for a system with both nanoparticles and
liquid. To evolve the liquid, a random site on the lattice is picked. The
MC algorithm described above is used, however any move involving a
nanoparticle is forbidden during a liquid step.

The nanoparticles are modelled differently. Instead of selecting any random
particle from the system, we keep an explicit list of nanoparticle
positions and select a nanoparticle from this list to attempt the following
dynamics: Firstly, to prevent nanoparticles from being left
floating when the surrounding liquid moves away, after selecting a
nanoparticle, we first check if there is a vacancy in the lattice site below the
chosen nanoparticle. If there is, the nanoparticle is moved down to that
empty site, finishing the move. If the site below is non-empty or is part of
the surface, we then perform a weighted sum over
the neighbouring lattice sites to determine how much liquid there is in
the vicinity of the nanoparticle. We calculate the quantity
$\bar{l}_{\i}=\sum_{\j}c_{\i\j} l_{\j}$, where the coefficients $c_{\i\j}$ are
the same as those used to calculate the energy, given in Eq.\ \eqref{eq:c_ij}.
If $\bar{l}_{\i}<5/3$ then the nanoparticle move is rejected. Only if
$\bar{l}_{\i}\geq5/3$ we do allow the nanoparticle to move, swapping
with one of the neighbouring liquid particles, as per steps 2--4 above.
This is done to prevent excessive nanoparticle movement once most of the
liquid has evaporated, since the physical origin of the nanoparticle
dynamics is the Brownian motion due to being suspended in the liquid. If
there are not enough liquid particles neighbouring the nanoparticle, then
it remains stationary. The threshold value $5/3$ was determined empirically;
a lower value makes the nanoparticles too mobile on the dry surface, but
higher values leads to the formation of immobile nanoparticle clusters.
{ Note that the algorithm described above for evolving the
nanoparticles violates detailed balance. This is in keeping with previous
MC models for systems of this kind \cite{rabani2003, pauliac2008, vancea08} and is a consequence of the facilitated dynamics of the nanoparticles. Of course, for the liquid there is detailed balance.}

Liquid and nanoparticles evolve at different rates. We perform $M$
liquid steps for every nanoparticle step. This ratio determines the
diffusion coefficient of the nanoparticles in the liquid
\cite{rabani2003, pauliac2008, vancea08}. We set the value of $M$
to depend on
the ratio of nanoparticles to non-nanoparticles in the system as:
\begin{equation}
M = \xi \frac{V - N\sigma^3}{N\sigma^3}
\label{eq:M}
\end{equation}
where $V$ is the volume of the system and $N$ is the
total number of nanoparticles.
For all simulations in this paper, we use a value of $\xi = 0.2$. For
typical systems this corresponds to a value of $M \approx 30$.
Eq.~\eqref{eq:M} is required to prevent the nanoparticles ``speeding
up'' as the liquid evaporates from the system, which decreases the ratio
of liquid to nanoparticles.


\subsection*{Diffusion coefficient}\label{diffusion-coefficient}

In what follows the system is referred to as having evolved for
a time of $x$ Monte Carlo steps (MC steps), which means that there
has been an attempted move on average $x$ times per lattice
site. To relate MC steps to the physical time scales, the diffusion
coefficient of a single nanoparticle moving though the bulk liquid is
determined.

This is calculated by running multiple simulations with a single
nanoparticle in a system full of liquid. The distance $r$
that the nanoparticle travels is recorded at certain time
intervals. A plot of $\langle r^2 \rangle$ against the number of MC
steps is then made. Note that $\langle {\cal P}\rangle$ denotes the
statistical average of any quantity ${\cal P}$. Using the
relation~\cite{einstein1905}
\begin{equation}
  \langle r^2 \rangle  = 6Dt
\end{equation}
where $D$ is the diffusion coefficient and $t$ is time, the value of $D$
can be determined from the gradient of the plot.

For a system with $\mu/\epsilon_{ll}=6$ (a system filled with liquid),
$k_BT/\epsilon_{ll}=0.6$ and  averaging over $10,000$ simulations, a value of
$D=2.6\times 10^{-4}\,\sigma^2\, {\text{MC step}}^{-1}$ was found.
Thus, the Brownian timescale $\tau_B\equiv\sigma^2/6D=650$ MC
steps. $\tau_B$ is the time it takes on average for a nanoparticle to
diffuse a distance of order its own diameter. We obtain a similar value at
the higher temperature $k_BT/\epsilon_{ll}=1.0$, since the value of $D$
only starts to change when the temperature is high enough or the chemical
potential is low enough that the
density of the vacancies in the liquid becomes sizeable.
Although we specify times below in units of MC steps, knowing the
value of $D$ allows to easily relate to the true timescales in a given
system.


\subsection*{Determination of contact angles}\label{contact-angles}

\begin{figure}[t]
\centering
\includegraphics{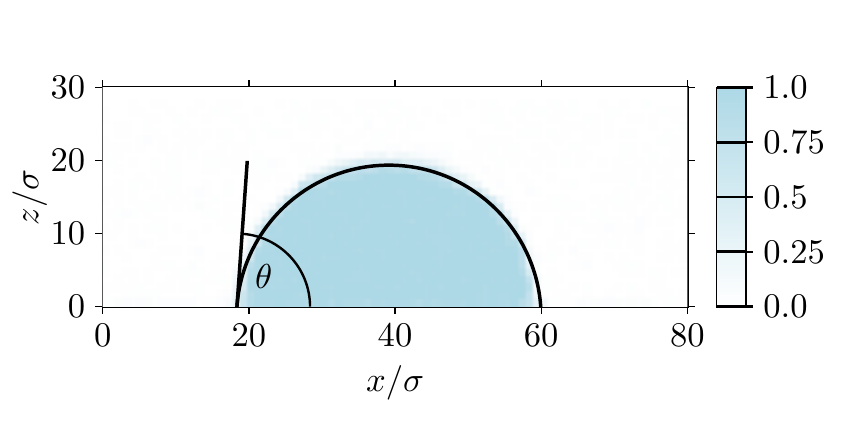}
\caption{%
  A density profile for a drop at equilibrium with $k_{B}T/\e_{ll}=1$
  and $\e_{wl}/\e_{ll}=0.7$ obtained by averaging along the length of
  the liquid drop. The approximating circle used to estimate the contact
  angle is shown as the black line. This circle is calculated using the
  Taubin circle fitting method \cite{taubin} on the boundary points of
  the profile.
}
\label{fig:circle-fit}
\end{figure}

Once the system, such as that illustrated in Fig.~\ref{fig:initial}, has
reached equilibrium we can measure the contact angle. This is done by
taking an average along the length of the liquid ridge in the $y$-direction
(c.f.~Fig.~\ref{fig:initial}). We average over the configurations
of a liquid ridge instead of a hemispherical drop
because this is easier to measure and gives us more samples
to average over. This average calculates a density profile $\rho_\i =
\langle l_i \rangle$. From this density profile, we define the liquid
drop to be where $\rho_\i\sigma^3 > 0.5$. We then fit a circle to the
top portion of the boundary of the drop using the Taubin circle fitting
method~\cite{taubin}, illustrated in Fig.~\ref{fig:circle-fit}. From
this circle, it is then straightforward to determine the contact angle,
which is the angle made with the surface. The density profile in
Fig.~\ref{fig:circle-fit} is for a system with temperature
$k_{B}T/\e_{ll}=1.0$ and wall attraction strength $\e_{wl}/\e_{ll}=0.7$.
The 0.7 value corresponds to a weakly hydrophilic interface and so the
liquid does not spread and forms a drop with a contact angle
$\approx75^\circ$. Increasing $\e_{wl}$ decreases the contact angle,
corresponding to the surface becoming more hydrophilic. On the other
hand, decreasing $\e_{wl}$ makes the surface hydrophobic.

\begin{figure}[t]
\centering
\includegraphics{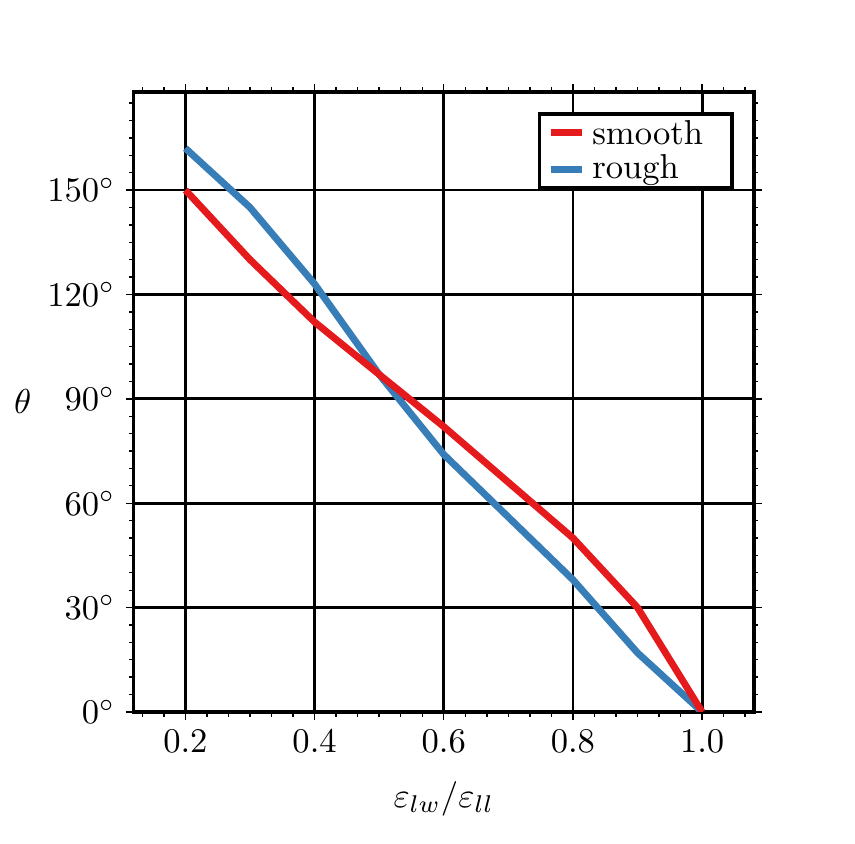}
\caption{%
  The contact angle $\theta$ plotted as a function of the surface
  attraction $\e_{wl}$ with $k_BT/\e_{ll} = 1.0$, for both a smooth and a
  rough surface. We see that increasing
  the attraction due to the surface decreases the contact angle. When
  $\e_{wl} > \e_{ll}$, the drops wet the surface.
}
\label{fig:comp-angles}
\end{figure}

In Fig.~\ref{fig:comp-angles} we display a plot of the contact angle as
the surface-liquid attraction strength $\e_{wl}$ is varied, for the
temperature $k_BT/\e_{ll}=1.0$. { In addition to results for this
`smooth' surface, we also include in
Fig.~\ref{fig:comp-angles} results for a `rough' surface,
discussed further below.} We see that as the attraction strength
increases, the contact angle decreases, until eventually at $\e_{wl}
\approx \e_{ll}$ there is a wetting transition to a state where the
liquid wets the surface, with contact angle $0^\circ$. For small values
of $\e_{wl}$ the surface only weakly attracts the liquid, corresponding
to a strongly hydrophobic surface on which the drop takes a shape that
is close to a full circle, with a large contact angle. Owing to the way
we define the wall potential, the contact angle plot in
Fig.~\ref{fig:comp-angles} varies only weakly with the temperature in
the range $0.6 < k_BT/\e_{ll} < 1.2$, the range in which most of our
results are calculated. At higher temperatures, one should expect the
wall attraction strength for wetting  to be lower. However, at higher
temperatures the interfacial fluctuations become significant and the
system is no longer in the regime relevant to modelling the drying of
inkjet printed drops. At lower temperatures (results not displayed), the
simulations become slow and the system becomes hard to equilibrate.

We also display in Fig.~\ref{fig:comp-angles} the contact angle obtained
for the liquid on a rough surface. This surface is
physically rough on the scale of the lattice, modelled by randomly raising
and lowering respectively one third of the blocks on the surface by one
lattice spacing $\sigma$. When the wall is sufficiently attractive, for
$\e_{wl}/\e_{ll}>0.5$, this generates a surface that contains many
pits, into which liquid condenses (from the vapour) and becomes
trapped. This makes the surface effectively more attractive
and so the contact angle in this regime is decreased, compared to
the smooth surface. However, for $\e_{wl}/\e_{ll}<0.5$ the surface
roughness makes the surface more hydrophobic and with a larger
contact angle than the smooth surface
with the corresponding value of $\e_{wl}$. This is the well-known
lotus effect used to create superhydrophobic surfaces via surface roughness
\cite{wenzel1936resistance, cassie1944wettability, bico1999pearl, de2004capillarity}.


\section{Bulk solvent phase behaviour}
\label{sec:bulk}

Understanding the behaviour of the liquid in equilibrium gives us
insight into how the liquid behaves out of equilibrium. Calculating the
binodal allows us to pick parameters that correspond to a suitably high
density liquid phase coexisting with low density vapour phase.

The binodal gives the coexisting density values for a system in
equilibrium. Two coexisting phases have the same chemical potential,
temperature and pressure in each phase.

Since we do not need to know the binodal densities with great accuracy
we calculate the binodal by performing simulations in a long, narrow box
of size $10\sigma\times10\sigma\times80\sigma$, with periodic boundary
conditions, treated in the canonical ensemble. Initially one end of the
box of filled with liquid particles, with the other half being empty.
The simulation then equilibrates in a state with half the system in the
liquid phase, coexisting with the other half containing the vapour.

\begin{figure}[t]
\centering
\includegraphics{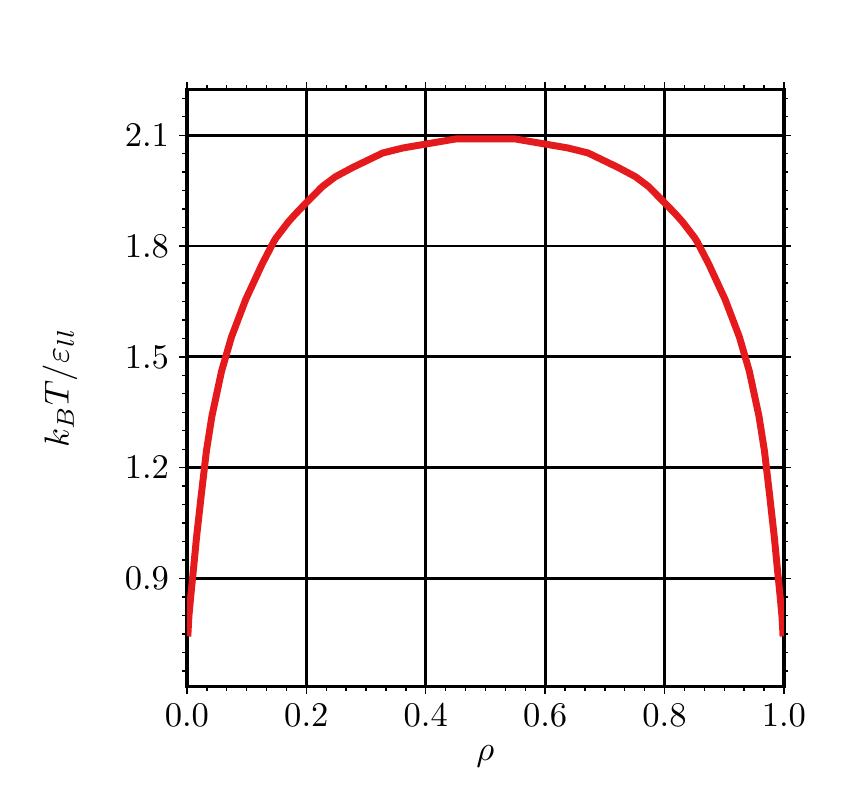}
\caption{%
  The bulk fluid binodal, which gives the densities of the coexisting
  gas and liquid phases as the temperature is varied.
}
\label{fig:binodal}
\end{figure}

To estimate the density of the two coexisting phases we first calculate the mean
density $\rho_\i$ of each $10\sigma\times10\sigma$ layer of the box. The
layer densities are then split into two groups: those with $\rho_\i>0.5$
and those with $\rho_\i<0.5$. In each of these groups the statistical
outliers are eliminated, since these are layers that correspond to the
interface between the gas and the liquid. Then the mean of the remainder
in each group is used as the density on the binodal. The result of this
approach, over a range of temperatures, yields the binodal displayed in
Fig~\ref{fig:binodal}. For example, when
$k_BT/\varepsilon_{ll}=0.9$ the density of
the coexisting liquid and vapour is $\rho_l=0.99$ and $\rho_g=0.01$.

In the vicinity of the critical temperature $T_c$, the binodal can not
be calculated with any great accuracy using the approach described
above, due to the fact that the average densities of the two coexisting
phases become rather similar and also because the system is strongly
fluctuating with a diverging correlation length at $T_c$~\cite{hansen,
landau2014guide}. The top of the binodal curve displayed in
Fig.~\ref{fig:binodal} was estimated by inspecting the histogram of
densities $\rho_\i$ in each layer. Below the critical temperature, this
histogram has two distinct maxima, corresponding to the two phases. We
use the density value at each of these maxima as our estimate for the
densities of the two coexisting phases. For $T>T_c$, the density
histogram has only one maximum, at $\rho_\i=0.5$. Based on this method
we find that the critical temperature is $k_BT_c/\e_{ll}=2.08\pm 0.02$.


\section{Evaporating droplets}
\label{sec:droplets}

For evaporation to occur, the statistical mechanics of the system must
be done in the semi-grand canonical ensemble: the liquid is treated
grand canonically, whilst the nanoparticles are dealt with canonically.
The liquid is treated grand canonically because the vapour above the
drop acts as a reservoir with chemical potential $\mu$, with which the
system can exchange particles, allowing the number of liquid particles
in the system to vary over time.
This is achieved by periodically setting the density of the top layer of
the system to the low density result $\rho_\i \approx e^{\beta\mu} / (1
+ e^{\beta\mu})$~\cite{hughes2014, hughes2015}.
This enables the removal of particles from the system as the liquid drop
evaporates. In contrast, the nanoparticles are treated canonically,
since the number of nanoparticles in the system is fixed over time.
In contrast, as discussed above,
in order to determine the contact angle of a drop of liquid on
the surface, we must treat it canonically, with a fixed number of liquid
particles in the system.

\begin{figure}[t]
\includegraphics{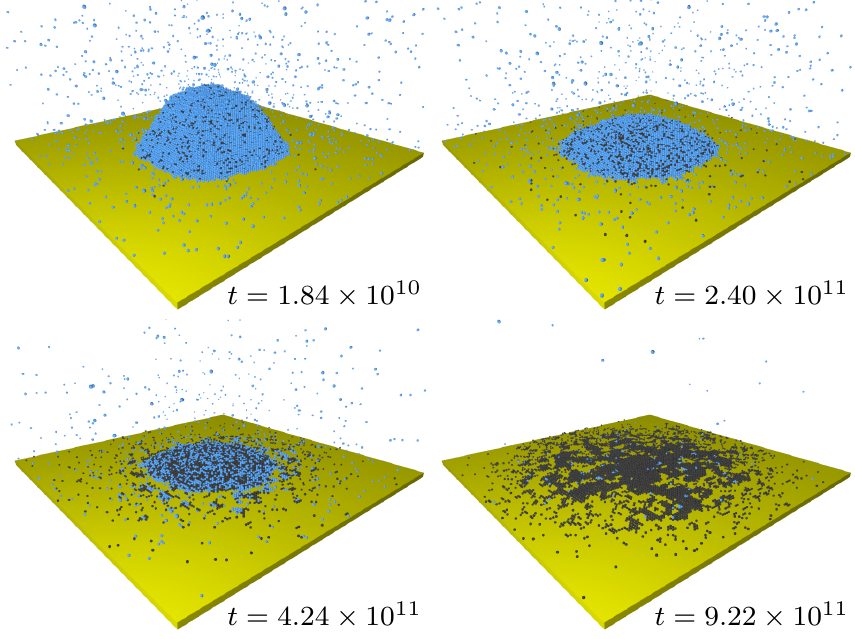}
\caption{%
  Droplet evaporation simulation, for $k_BT/\e_{ll}=0.8$,
  $\mu/\e_{ll}=-9$, $\phi=0.15$, $\e_{nn}/\e_{ll} =1.5$,
  $\e_{nl}/\e_{ll}=1.25$, $\e_{lw}/\e_{ll}=\e_{nw}/\e_{ll}=0.8$
  in a system with surface area
  $120\sigma\times120\sigma$ and box height $80\sigma$. Times, in
  units of MC steps, are
  given at the bottom right of each snapshot. After most of the liquid has
  evaporated, the nanoparticles diffuse out over the smooth surface,
  with dynamics facilitated by the small, but non-zero, vapour density.
\label{fig:evap-smooth}
}
\end{figure}

We initiate the system with a fraction $\phi$ of the liquid particles
replaced by nanoparticles.
Fig.~\ref{fig:evap-smooth} shows the evaporation of a droplet containing
nanoparticles with initial concentration $\phi=0.15$. The substrate area is
$120\sigma\times120\sigma$ and the height of the top of
the simulation box above the substrate is $80\sigma$. The
initial droplet consists of a hemisphere with a radius of $40\sigma$
with the vertical part linearly scaled to have a height of $24\sigma$.
The chemical potential is $\mu/\e_{ll}=-9$ and temperature
$k_BT/\e_{ll}=0.8$ which corresponds to an equilibrium vapour with a
density $\rho_g=0.001$. The interaction parameters are
$\e_{lw}/\e_{ll}=\e_{nw}/\e_{ll}=0.8$, $\e_{ln}/\e_{ll}=1.25$
and $\e_{nn}/\e_{ll}=1.5$.

The drop in Fig.~\ref{fig:evap-smooth} initially spreads to cover a greater
area of the surface, since the starting configuration does not have
the equilibrium contact angle. However, over time, liquid evaporates and
the drop reduces in volume and so subsequently the area of the surface
covered by the drop decreases -- i.e.\ the contact line initially
advances and then later recedes. Owing to the smooth surface, the drop
retains a dynamic contact angle that is close in value to the
equilibrium contact angle throughout most of the time evolution. Since
the nanoparticles are attracted to the liquid they generally follow the
liquid.

After most of the liquid has evaporated there is then a further spreading
of nanoparticles over the surface. Because of the smoothness of the
surface and the non-zero vapour density, the residual liquid facilitates
a diffusive dynamics that allows the nanoparticles to spread out
over the surface to a state where
the average distance of the nanoparticles from
the centre of the system is larger than when the liquid is present.

\subsection*{Surface roughness}

The roughness of surfaces is known to play an important role in how
liquids spread. Surface roughness can hinder contact
line motion over the surface and can lead to significant differences
between the advancing, receding and
equilibrium contact angles~\cite{de2004capillarity}. We consider two
different methods of modelling the effect of surface roughness. The
first is to introduce a dynamic rule that forbids moves parallel to the
surface for all particles in contact with the surface. This is equivalent to
a no-slip boundary condition. Thus, for a contact line to advance,
particles in the second layer of lattice sites or higher
above the surface must advance and then
drop down to wet the dry surface ahead of the spreading droplet. 

\begin{figure}[t]
\includegraphics{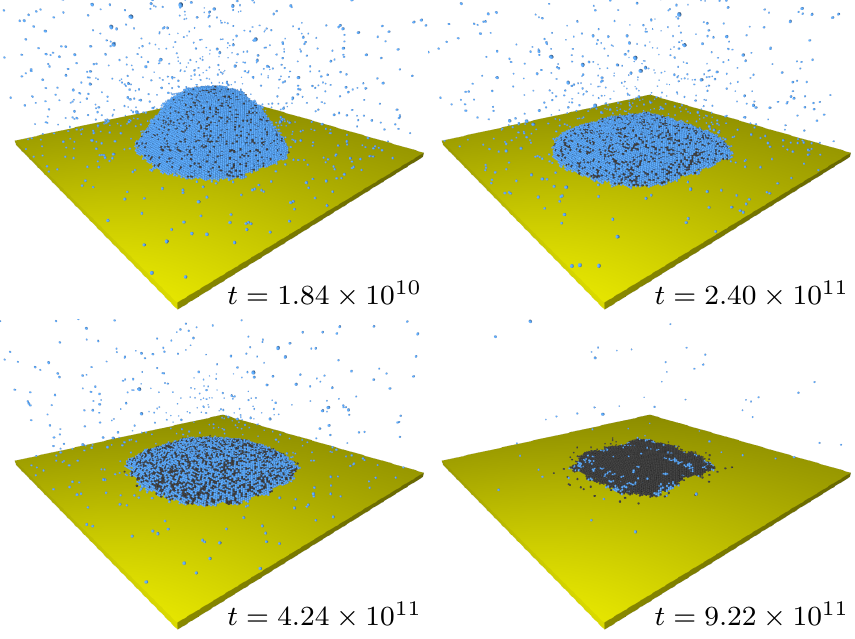}
\caption{%
  Snapshots of a liquid drop evaporating from a rough surface, with surface
  roughness modelled by incorporating a no-slip dynamic rule preventing
  motion at the surface being parallel to the surface.
  These are for the same times and parameter values
  as the smooth surface results in Fig.~\ref{fig:evap-smooth}.
\label{fig:evap-rule}
}
\end{figure}

Fig.~\ref{fig:evap-rule} shows snapshots from a simulations with the
same parameter values as the evaporation simulation in
Fig.~\ref{fig:evap-smooth} but with the no-slip dynamical rule forbidding
moves across the surface. The droplet still spreads
to a cover an area similar to that in the case with the smooth surface
-- i.e.\ to a value similar to that dictated by the equilibrium contact
angle for this particular value of $\e_{wl}$. We then find that
once most of the liquid has evaporated, the nanoparticles are left in an
almost uniform circle which has a slightly larger radius than the original
drop. There is also no further spreading out over the surface, even
though the vapour density is still non-zero.

\begin{figure}[t]
\includegraphics{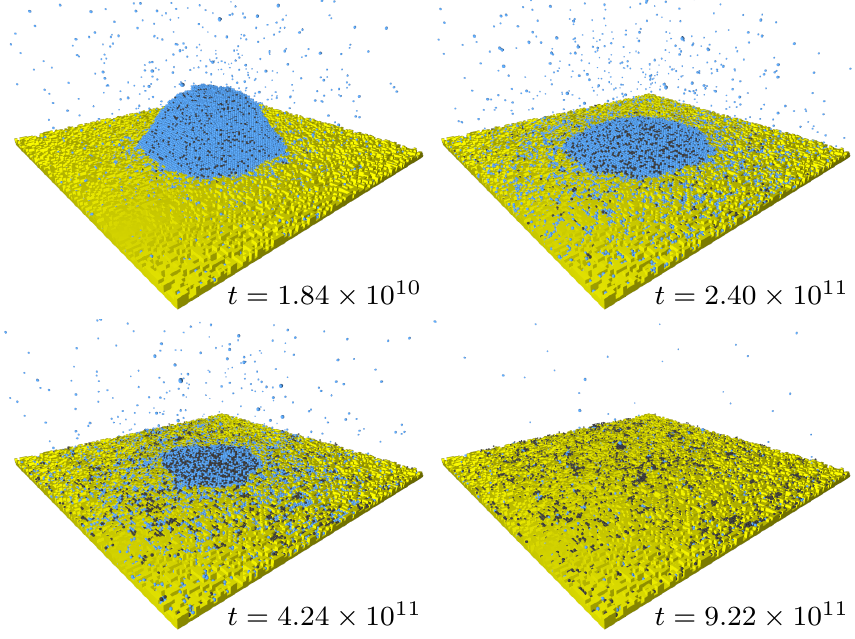}
\caption{%
  Snapshots of a liquid drop evaporating from a rough surface, the
  effect of which is modelled by setting the height of the surface to 
  randomly be 0 or $\pm\sigma$, each with equal probability.
  These are for the same times and parameter values as the cases
  in \ref{fig:evap-smooth} and Figs.~\ref{fig:evap-rule}.
\label{fig:evap-rough}
}
\end{figure}

We have also investigated droplet evaporation from the rough surface
considered at the end of Sec.\ \ref{methodology} that is 
physically rough on the scale of the lattice, made by randomly setting the
hight of the surface to be 0 or $\pm\sigma$, each with equal probability.
Results for this surface are displayed in Fig.~\ref{fig:evap-rough}.
Recall that for $\e_{wl}/\e_{ll}>0.5$ the contact angle is less than on
the corresponding flat surface (see Fig.\ \ref{fig:comp-angles}).
This second approach to
modelling surface roughness generates a wall that contains many
pits, within which liquid becomes trapped. This leads to a much higher
amount of liquid remaining adsorbed on the surface than in the cases
in Figs.~\ref{fig:evap-smooth} and \ref{fig:evap-rule}. The adsorbed
liquid facilitates the spreading of the
nanoparticles over the surface out to distances well beyond where
the liquid droplet was located. Whilst this facilitated dynamics is
interesting, it is not what is observed on the experimental surfaces
of interest here.

\begin{figure}[t]
\centering
\begin{tabular}{c c c}
  \includegraphics{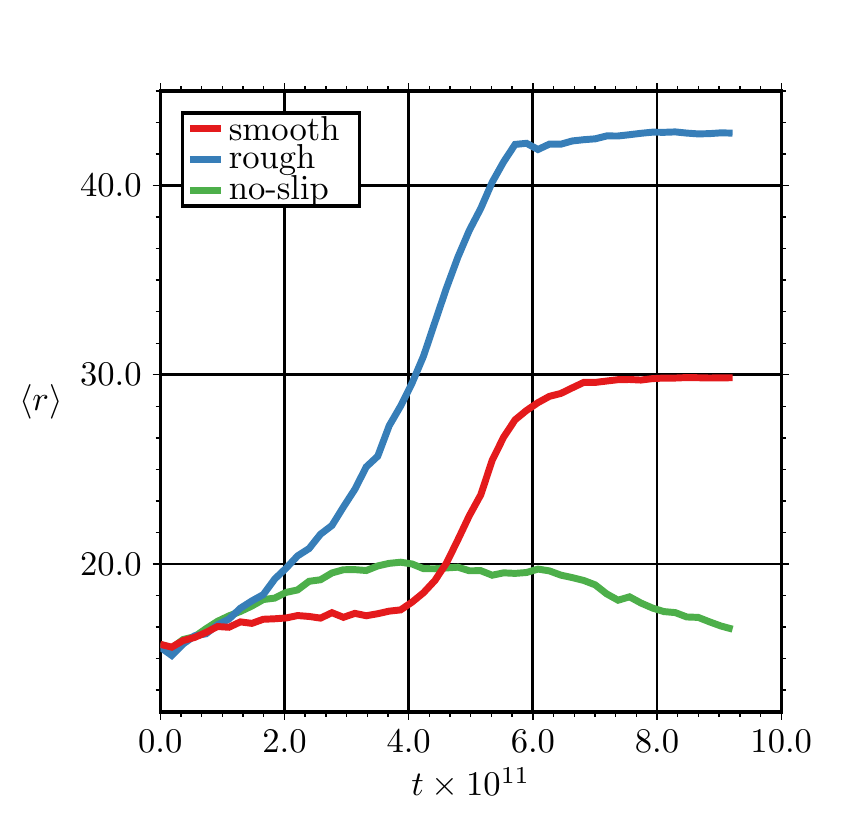}
\end{tabular}
\caption{%
  Plot of the average distance of the nanoparticles from the centre over
  time, for the three cases of (i) a smooth surface
  [Fig.~\ref{fig:evap-smooth}], (ii) a rough surface where the effect of the
  surface roughness is modelled by a no-slip dynamic rule 
  [Fig.~\ref{fig:evap-rule}] and (iii) surface roughness modelled by making
  the surface height randomly higher/lower than the average
  [Fig.~\ref{fig:evap-rough}].
}
\label{fig:nano-radius}
\end{figure}

In Fig.~\ref{fig:nano-radius} we display a plot of the mean distance
$\langle r\rangle$ of the nanoparticles from the centre of the system
(the centre of where the droplet was initiated) as
a function of time for the three different surface roughness cases.
Initially the average radius increases due to the droplet spreading in
order to try and reach the equilibrium contact angle. However, except in
the case where we model the surface roughness via the no-slip dynamic
rule, once the liquid has evaporated, $\langle r\rangle$ further
increases because the nanoparticles continue to spread out over the
surface, facilitated by the vapour of liquid particles. The plateau
value of $\langle r\rangle$ for the physically rough surface
(Fig.~\ref{fig:evap-rough}) is even greater than the smooth surface
case (Fig.~\ref{fig:evap-smooth}) due to the higher amount of
liquid adsorbed on the surface, in the surface pits. Eventually,
$\langle r\rangle$ tends to a constant value as the spreading
nanoparticles become trapped in the pits in
the surface. For the case with the no-slip dynamic rule modelling
surface roughness,  as the droplet spreads and evaporates,
$\langle r\rangle$ reaches a maximal value at around $t=4\times 10^{11}$.
It then decreases slightly as the droplet contact line starts to recede,
due to the droplet volume being decreased by the evaporation.

The results in Figs.~\ref{fig:evap-rule} and \ref{fig:nano-radius} show that
incorporating the effects of surface roughness via the no-slip dynamical
rule seems to model the required physics. It also has the additional advantage of
not introducing an additional length scale to be considered, namely the
length scale of the surface roughness. Thus, this
is the model we adopt henceforth to model the effects of surface roughness.


\section{Modelling the ink drying process}\label{nanoparticles}

\subsection*{Evaporating liquid bridge over a hydrophobic strip}

\begin{figure*}
  \includegraphics{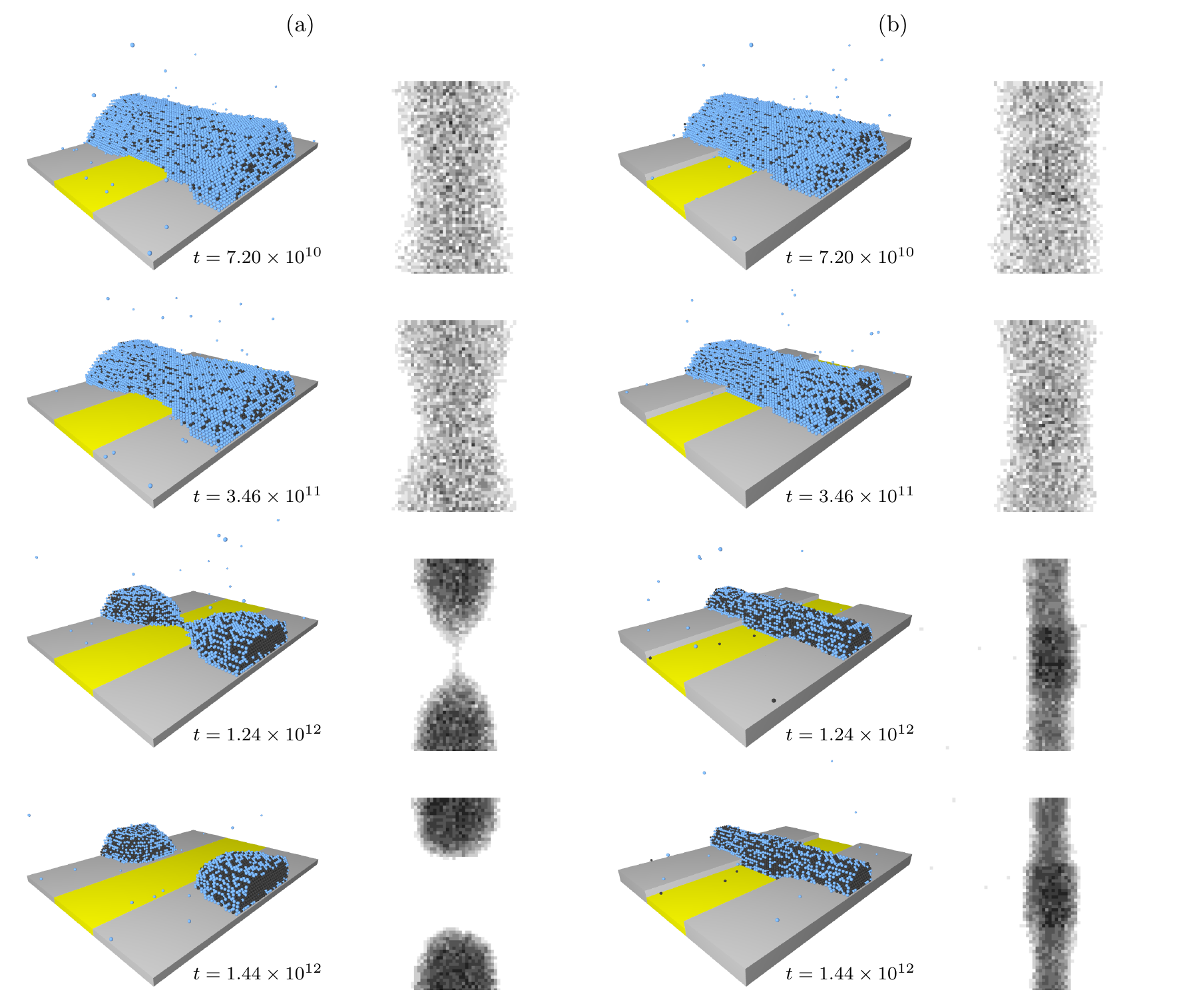}
\caption{%
  Time series from the drying of the liquid from the surface, when
  $\e_{nl}/\e_{ll} = 1.5$, $\e_{nn}/\e_{ll}=2$, $\phi=0.2$,
  $k_{B}T/\e_{ll}=0.6$ and $\mu/\e_{ll}=-6$. The attraction strength
  with the hydrophobic yellow part A surface is $\e^{\rm
  A}_{wl}/\e_{ll}=\e^{\rm A}_{wn}/\e_{ll}=0.4$, while the interaction
  with the grey part B strips either side has strength $\e^{\rm
  B}_{wl}/\e_{ll}=\e^{\rm B}_{wn}/\e_{ll}=0.8$. On the right of each
  snapshot is the nanoparticles density distribution for that
  snapshot as viewed from above. The results on the left are for the
  case when there is
  no step ($h=0$) going from the part B to part A. In this case, as the
  liquid evaporates, it also dewets from the surface, breaking the
  bridge. The results on the right correspond to when there is a step of
  height $h=2\sigma$. This step prevents the dewetting, so that as the liquid
  evaporates, the nanoparticles gather to form a bridge. The times $t$ are given
  in terms of average number of MC steps per lattice site.
}
\label{fig:strip-smooth-a}
\end{figure*}

In Fig.~\ref{fig:strip-smooth-a} we display snapshots as the liquid
evaporates from a surface containing a hydrophobic strip. The
interaction parameters are given in the figure caption. The surface is
smooth -- i.e.\ we do not implement the no-slip dynamical rule. The
chemical potential in the vapour is set to be $\mu/\e_{ll}=-6$, which
corresponds to the vapour phase being the thermodynamic equilibrium
state, so the liquid seeks to evaporate from the surface. The substrate
is made of a central hydrophobic strip of width 20$\sigma$ (coloured
yellow) which we denote region A, with $\e^{\rm A}_{wl}/\e_{ll}=\e^{\rm
A}_{wn}/\e_{ll}=0.4$, i.e.~only weakly attracting the liquid and the
nanoparticles. From Fig.~\ref{fig:comp-angles} we see that on this part
of the surface the liquid has contact angle $\theta\approx110^\circ$.
Either side of this (coloured grey) is region B, where
the surface is hydrophilic, having attraction strength parameters
$\e^{\rm B}_{wl}/\e_{ll}=\e^{\rm B}_{wn}/\e_{ll}=0.8$, corresponding to
$\theta\approx60^\circ$. When there is no
step in height from region B to region A ($h=0$), then
Fig.~\ref{fig:strip-smooth-a} shows that during the drying, the liquid
dewets from the hydrophobic part of the surface, breaking the
nanoparticle bridge at time $t\approx1.2\times10^{12}$ MC steps.
When there is a small step of height $h=\sigma$ (results not displayed),
then the behaviour
is similar, although the breaking of the bridge is slightly delayed.
In contrast, a step of height $h\geq2\sigma$
enables the liquid bridge to remain intact as it dries, so that a
nanoparticle bridge is formed, spanning the hydrophobic part of the
surface. The nanoparticle density is even
slightly increased on the hydrophobic part of the surface when
$h\geq2\sigma$ (see Fig.~\ref{fig:strip-smooth-a}).

The reason a step enables the liquid bridge to remain is that a corner
is created into which the liquid is strongly attracted. The ability of
corners and wedges to promote wetting by a liquid is well
known~\cite{concus1969, pomeau1986, hauge1992, rejmer1999, MP13a,
MP13b}. Since surface roughness can also modify the wetability of
surfaces, a combination of steps and roughness can be used to control
dewetting.

\begin{figure*}[t]
\centering
\includegraphics{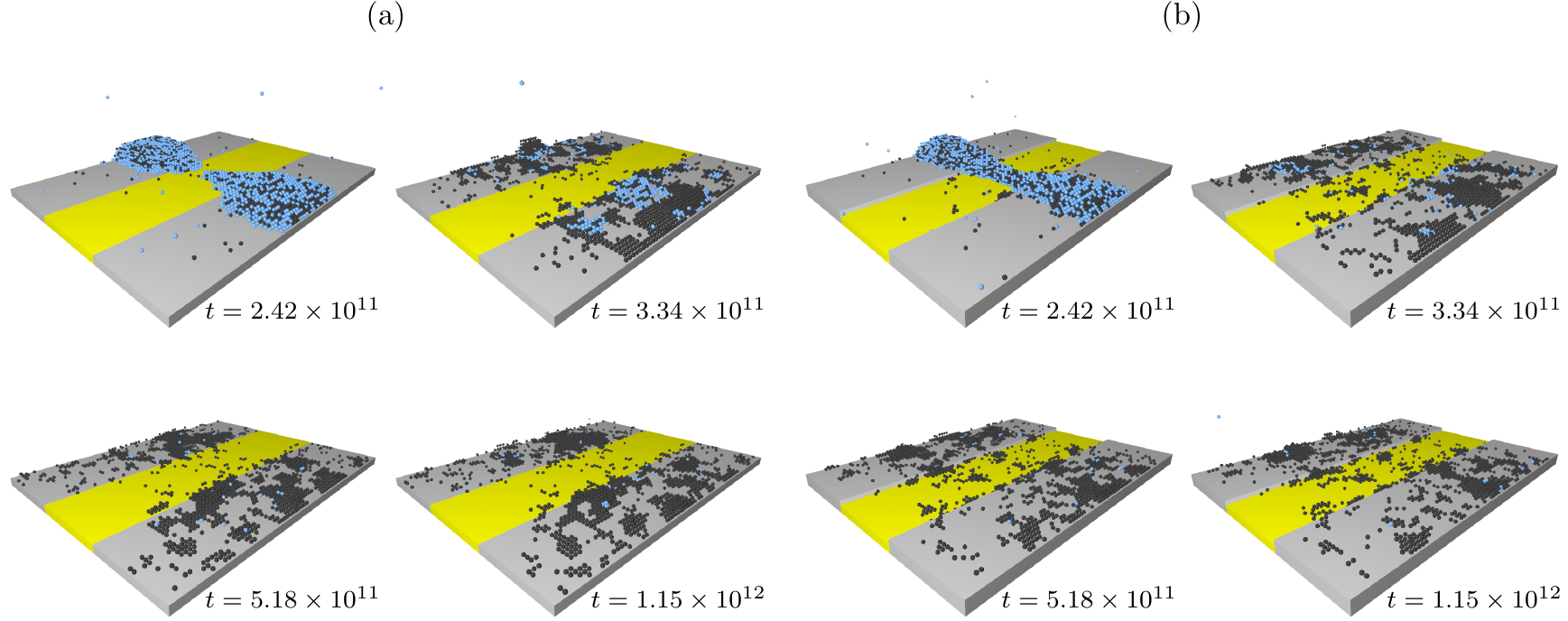}
\caption{%
  Snapshots from a simulation of a liquid bridge drying from the smooth surface
  with $\e_{nl}/\e_{ll} = 1.25$, $\e_{nn}/\e_{ll} = 1.5$,
  $k_BT/\e_{ll}=0.6$, $\phi=0.1$, $\mu/\e_{ll} = -6$. The attraction
  strength with the hydrophobic yellow part A surface is $\e^{\rm
  A}_{wl}/\e_{ll}=\e^{\rm A}_{wn}/\e_{ll}=0.4$, while the interaction
  with the grey part B strips either side has strength $\e^{\rm
  B}_{wl}/\e_{ll}=\e^{\rm B}_{wn}/\e_{ll}=0.8$. In the four snapshots on
  the left in (a) there is no difference in height between the two
  surfaces ($h=0$). In the four on the right (b) the hydrophilic part B
  (in grey) is raised a distance $h=\sigma$ above part A.
}
\label{fig:strip-smooth-b}
\end{figure*}

Fig.~\ref{fig:strip-smooth-b} shows results from a case when the
nanoparticles are less strongly attracted to one another, which enhances
the spreading over the hydrophilic part of this (smooth) surface,
compared to the case in Fig.~\ref{fig:strip-smooth-a}. With no step
present ($h=0$), the bridge of liquid breaks at the time $t\approx2.4 \times
10^{11}$ MC steps and the nanoparticles temporally group together with
the remaining liquid, but eventually spread out over the hydrophilic
region. With a step of height $h=\sigma$, the bridge still breaks at
$t\approx2.4 \times 10^{11}$ MC steps. More nanoparticles remain at the corner
formed from the step but the end result is similar to the case with no
step. Although not shown here, when $h=2\sigma$, the connection breaks
at $t\approx2.1 \times 10^{11}$ MC steps but the break occurs in the hydrophilic
region and the nanoparticles collect in the hydrophobic region,
scattering randomly as the rest of the liquid evaporates, due to the
smoothness of the surface.

\begin{figure*}
\includegraphics{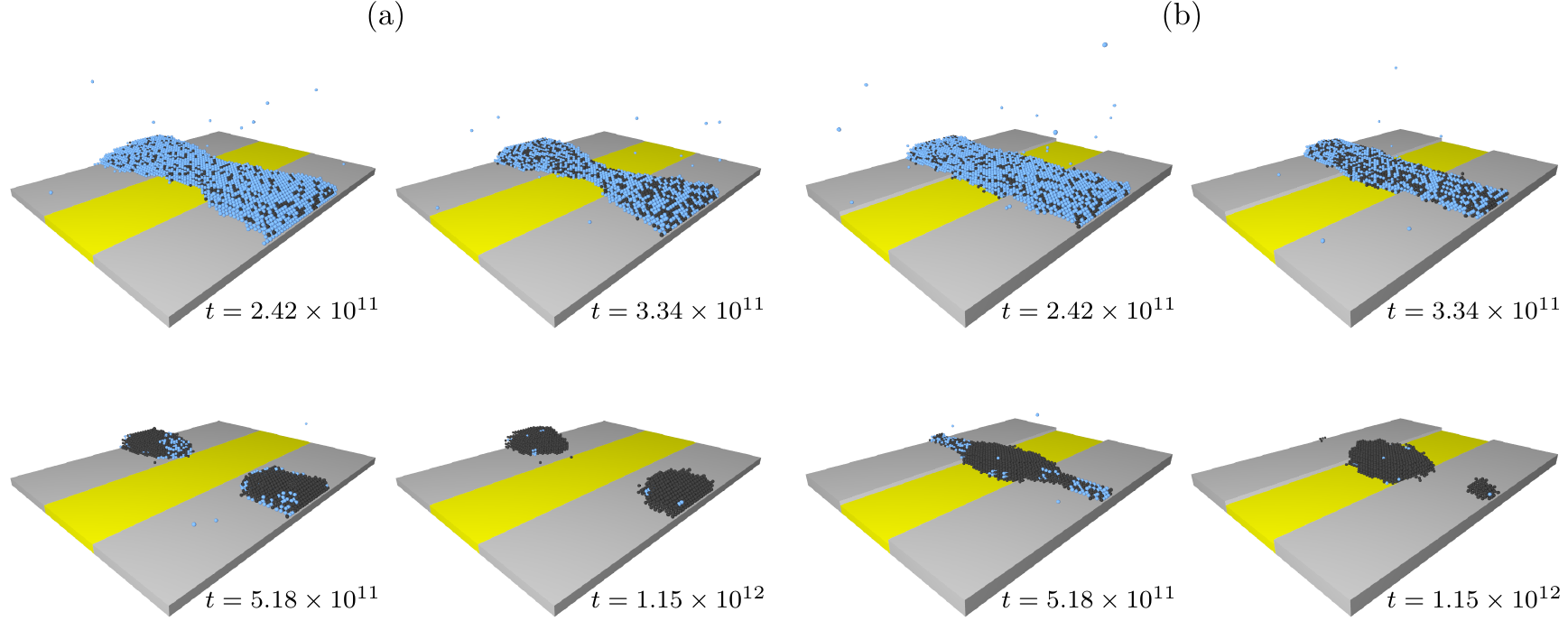}
\caption{%
  Time series from a bridge of liquid drying from a rough surface
  (modelled using the no-slip dynamical rule), with the same parameter
  values as given in caption of Fig.~\ref{fig:strip-smooth-b}. The four
  on the left (a), are snapshots for the case when there is no step
  ($h=0$) in the height of the surface. In this case, as the liquid
  evaporates, it also dewets from the surface, breaking the bridge. The
  four on the right (b) correspond to a step of height $h=\sigma$ between
  the hydrophobic and hydrophilic parts of the surface. The times $t$ are given in
  terms of average number of MC steps per lattice site, taken at the
  same times as in the previous figure.
}
\label{fig:strip-rule}
\end{figure*}

Fig.~\ref{fig:strip-rule} shows results from a simulation where the
parameters are the same as in Fig.~\ref{fig:strip-smooth-b}, except here
we assume the surface is rough, i.e.\ we implement the no-slip dynamical
rule. When there is no step ($h=0$), the bridge breaks at $t\approx4.4 \times
10^{11}$ MC steps. When the step height $h=\sigma$, the bridge connection
almost holds, but eventually breaks at $t\approx5.2 \times 10^{11}$ MC steps.
Interestingly, however, due to the attractive step from the hydrophobic
to the hydrophilic region, most of the nanoparticles are stabilised in a
cluster on the hydrophobic region.

\subsection*{Evaporating film over an hydrophobic strip}

\begin{figure*}[t]
\centering
\includegraphics{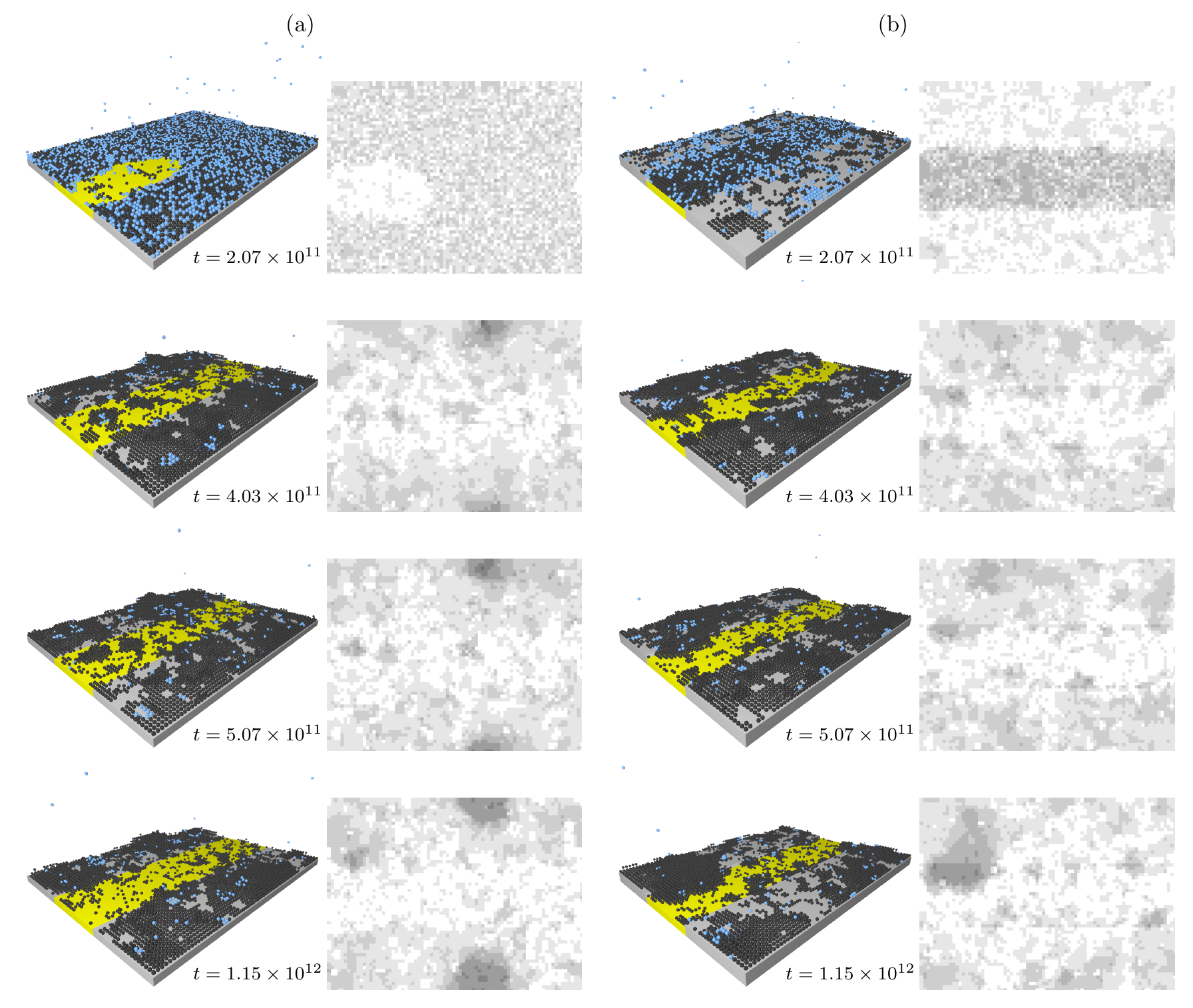}
\caption{%
  Snapshots of a uniform film of nanoparticle suspension drying from a
  smooth surface. On the right of each is a grey-scale
  density profile of the nanoparticles viewed from above.
  The parameter values are the same as in Fig.~\ref{fig:strip-smooth-b}
  except $k_BT/\e_{ll}=0.76$. The results on the left in (a) are for $h=0$
  and those on the right in (b) are with $h=\sigma$. 
}
\label{fig:block-smooth}
\end{figure*}

\begin{figure*}[t]
\includegraphics{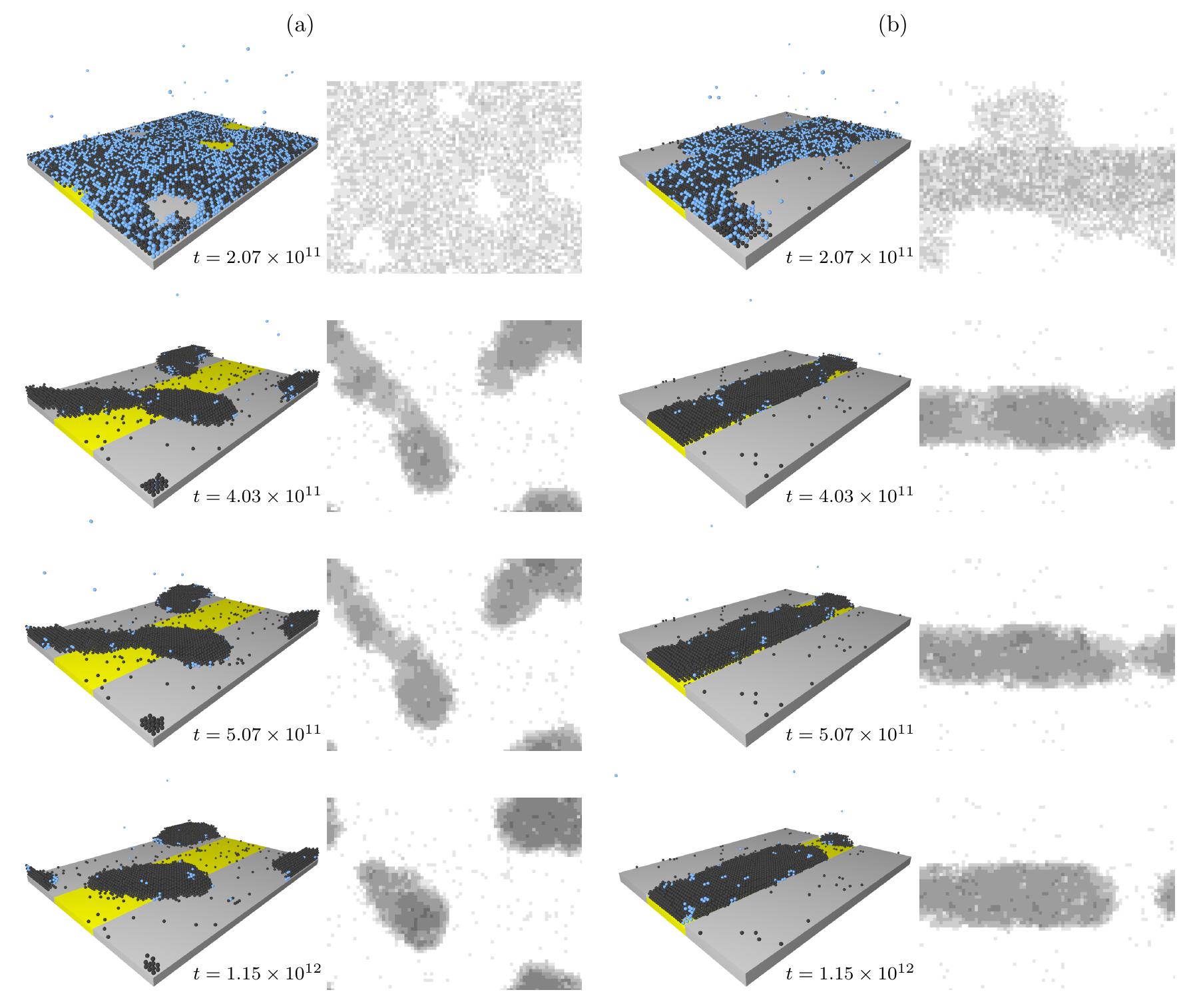}
\caption{%
  Snapshots of a uniform film of nanoparticle suspension drying from a
  rough surface, modelled via the no-slip dynamic rule. On the right of each
  is a grey-scale density profile of the nanoparticles viewed from above.
  The parameter values are the same as in Fig.~\ref{fig:block-smooth}.
  The results on the left in (a) are for $h=0$
  and those on the right in (b) are with $h=\sigma$.
}
\label{fig:block-rule}
\end{figure*}

Figs.~\ref{fig:block-smooth}~and~\ref{fig:block-rule} illustrate the
situation when a film of nanoparticle suspension that initially has
uniform thickness evaporatively dewets from the same surface considered
already, i.e.\ with both hydrophobic
and hydrophilic parts. All the parameters for the simulations in
Figs.~\ref{fig:block-smooth}~and~\ref{fig:block-rule} are the same as in
Fig.~\ref{fig:strip-smooth-b} except for the temperature which is
increased from $k_BT/\e_{ll}=0.6$ to $k_BT/\e_{ll}=0.76$, which slightly
speeds up the simulations.

Fig.~\ref{fig:block-smooth} for the smooth surface shows there are
differences between $h=0$ and $h=\sigma$. As the liquid
evaporates, in both cases holes appear in the film during the drying, at around
$t\approx2.0 \times 10^{11}$ MC steps. We are not able to
determine conclusively whether these holes are nucleated or are formed
via spinodal dewetting, which is expected to occur when the film thickness decrease
below a critical value \cite{kalliadasis2007, thiele2009, robbins2011}. In the
$h=0$ case, the holes appear first in the hydrophobic region. This leads to a dewetting
of the liquid from off the hydrophobic region, moving many of the nanoparticles
onto the hydrophilic region. In contrast, for the
$h=\sigma$ case, since the film is thicker over the hydrophobic
region, the holes instead appear first over the hydrophilic region. Thus, in
the $h=\sigma$ case, initially the dewetting from the hydrophilic part of the surface
leads to a clear increase in the density of the
nanoparticles on the hydrophobic region. However, they then subsequently move
back onto the hydrophilic part of the surface as the
evaporation continues. In both cases, after most of the liquid has evaporated,
the nanoparticles are distributed inhomogeneously over the surface, having
a greater density on the hydrophilic part of the surface. However, for the
$h=\sigma$ case, because the nanoparticles congregate at the corner
of the steps, there is therefore slightly more bare patches on
the hydrophilic part of the surface, compared to the $h=0$ case.

Fig.~\ref{fig:block-rule} shows snapshots from two simulations with the
no-slip dynamical rule, which prevents horizontal movement of particles
that are in contact with the surface. Holes in the film appear in a manner
similar to that observed in the early stages of the dynamics when the
surface is smooth (Fig.~\ref{fig:block-smooth}). However, once the holes
are formed, the dynamics is changed significantly. The surface
roughness results in the
nanoparticles becoming congregated in clumps and they spread far less
than in the case with the smooth surface. For the $h=0$ case in
Fig.~\ref{fig:block-rule}(a), the final state consists of the nanoparticles
being clustered into two
mounds with fewer lone nanoparticles than observed on the
smooth surface. Surprisingly, one of the nanoparticle clusters
spans the hydrophobic region of the surface. We believe this stems from
the interplay of the no-slip dynamics and the fact that the attraction of the
nanoparticles to each other is stronger than their attraction to the
surface.

In the case with a step of height $h=\sigma$ displayed in
Fig.~\ref{fig:block-rule}(b), the dewetting initiates in the thinner film on
the hydrophilic part of the surface, similar to the smooth surface case in
Fig.~\ref{fig:block-smooth}(b). This leads to the nanoparticles becoming
deposited on the hydrophobic region, similar to in the smooth surface
case, except that fewer nanoparticles remain on the hydrophilic region.
However, in contrast to the smooth surface case, ultimately the relative
lack of mobility leads to the nanoparticles remaining on the hydrophobic
region, forming a large cluster that is stabilised at the edges by the step.


\section{Conclusion}\label{conclusion}

In this paper we have presented a MC model of the drying of a
nanoparticle suspension on heterogeneous surfaces. This mixture
of liquid and nanoparticles is a simple model for the ink that is
used in the ink jet printing manufacturing process described in
\cite{walls}. The model contains parameters which can be
determined from experiments. Measuring the equilibrium
contact angle of drops of the liquid on the relevant surfaces, in conjunction
with the present work, allows the determination of the required values of the
liquid-liquid and liquid-wall attraction parameters. Similarly, knowledge
of the diffusion coefficient allows to relate the MC time step to the
experimental time scales. The model can include the effects of surface
roughness via a simple no-slip dynamical rule that forbids the
motion of all particles that are in contact with the surface.

A key finding of the present work is the observation that when printing a bridge over a
hydrophobic region to connect hydrophilic strips either side,
adhesion is improved when the hydrophobic strip is at a lower level than
the surrounding hydrophilic regions. We find that when the bridge does not
properly form, generally the break occurs over the hydrophobic strip. However,
for some parameter values occasionally the counter-intuitive result
occurs, whereby the film breaks, but with nanoparticles
congregating in the hydrophobic strip. This effect generally occurs when
considering the evaporation of a film of liquid, rather than a bridge.
That said, evaporating films can
still result in clumps of nanoparticles distributed over the two regions.

The results have shown the necessity to choose ink and surface
parameters carefully to obtain the best connections when ink jet
printing. { For example, it may be possible to enhance particle
bridge formation by adjusting the surface chemistry of the
nanoparticles to make them favour the hydrophobic portion of the surface.
This aspect has not been explored here. However, such enhancement might
also instead lead to results such as that in Fig.\ \ref{fig:strip-rule}, where
the bulk of the nanoparticles are deposited on the hydrophobic part
of the surface and the bridge is broken.}
Further work will directly relate the parameters in the model
Hamiltonian to the properties of the specialist materials used in the
printing process. Our work here shows that to fully understand the
observed phenomena
requires knowledge of both the fluid dynamics and the thermodynamics.

\section*{Acknowledgements}

\noindent The authors would like to thank Adam Brunton of M-SOLV for
useful discussions concerning ink jet printing and also Dmitri Tseluiko
for insightful comments on our work.


\end{document}